\documentclass[aps,epsf,amsmath,superscriptaddress,citeautoscript,twocolumn,
showpacs,floatfix]{revtex4}
\usepackage{bm}
\usepackage{amssymb}
\usepackage{graphicx}
\usepackage{amsmath, amsthm, amssymb, graphicx}
\usepackage[usenames]{color}

\usepackage{amsmath,amsthm,amsfonts,amscd,amssymb}
\usepackage{eucal} 	 	
\usepackage{verbatim}      	
\usepackage{makeidx}       	
\usepackage{braket}

\usepackage{natbib}
\usepackage{bibentry}
\usepackage{hyperref}

\newcommand{\be}{\begin{equation}}
\newcommand{\ee}{\end{equation}}

\begin{document}

\title{Correlated Worldline theory: Structure and Consistency}

\author{A.O. Barvinsky}
\affiliation{Theory Department, Lebedev Physics Institute, Leninsky Prospect 53, Moscow
119991, Russia}

\author{J. Wilson-Gerow}
 \affiliation{Pacific
Institute of Theoretical Physics, University of British Columbia,
6224 Agricultural Rd., Vancouver, B.C., Canada V6T 1Z1}
\affiliation{Department of Physics and Astronomy, University
of British Columbia, 6224 Agricultural Rd., Vancouver, B.C., Canada
V6T 1Z1}

\author{ P.C.E. Stamp}
 \affiliation{Pacific
Institute of Theoretical Physics, University of British Columbia,
6224 Agricultural Rd., Vancouver, B.C., Canada V6T 1Z1}
\affiliation{Department of Physics and Astronomy, University
of British Columbia, 6224 Agricultural Rd., Vancouver, B.C., Canada
V6T 1Z1}
\affiliation{Theoretical Astrophysics, Cahill,
California Institute of Technology, 1200 E. California Boulevard,
MC 350-17, Pasadena CA 91125, USA}
\date{{\small \today}}

\begin{abstract}
We give a formal treatment of the  "Correlated Worldline" theory of quantum gravity.
The generating functional is written as a product over multiple copies of the coupled
matter and gravitational fields; paths for fields are correlated via gravity itself. In the limit where the gravitational coupling $G \rightarrow 0$, conventional quantum field theory is recovered; in the classical limit $\hbar \rightarrow 0$, General Relativity is recovered. A formal loop expansion is derived, with all terms up to one-loop order $\sim O(l_P^2)$ given explicitly, where $l_P$ is the Planck length. We then derive the form of a perturbation expansion in $l_P^2$ around a background field, with the correlation functions given explicitly up to $\sim O(l_P^2)$. Finally, we explicitly demonstrate the on-shell gauge independence of the theory, to order $l_P^2$ in gravitational coupling and to all orders in matter loops, and derive the relevant Ward identities.
\end{abstract}


\maketitle


\section{ Introduction}
 \label{sec:intro}


\subsection{Background}
 \label{sec:background}

The effort to find a consistent theory of quantum gravity, which incorporates key
features of both quantum mechanics and General Relativity, has been going on now for many decades \cite{ashtekar74,bern02,carlip01,carlip15}. Roughly speaking, one can discern two points of view on how to do this:

On the one hand, one can assume that quantum mechanics (QM) is universally valid, and try to ``quantize" General Relativity (GR). Any problems that emerge - as they certainly do at very high energy - are then taken as a signal of the breakdown of QM in favour of some more fundamental theory, valid up to and beyond the Planck energy. This view is the most popular, and is assumed in, eg., string theory \cite{string}, loop quantum gravity \cite{ashtekar08}, or supersymmetric theories \cite{superS}. It is also often (but not always) assumed that at low energies, some effective quantum theory of gravity describes Nature.

Alternatively, one can argue that it is QM that should break down, even at low
energies, for sufficiently massive objects. Such arguments are very old, and have been
extensively reviewed \cite{jammer,zurekW}. They typically derive from the apparent
contradictions inherent in macroscopic quantum states \cite{AJL80}, with no particular
connection to gravity. Nevertheless, many authors, focusing on apparent contradictions
between QM and GR, have suggested that some sort of breakdown in QM might derive from
gravity; this qualitative idea also has a long history
\cite{einstein50,feynman57,kibbleb,penrose96}.

There are thus 2 diametrically opposed points of view here. In connection with the latter view - that we should look for a breakdown of QM - it is useful to emphasize several features of this question:

(i) some of the apparent contradictions between QM and GR are not restricted to high
energies. Thus, the infamous black hole information paradox \cite{hawking76}, so far
unresolved \cite{marolf17,unruh17}, exists already at energies well below the Planck
scale; and the problems associated with superpositions of different metric fields
\cite{kibbleb,penrose96,wald84} are manifested at any energy.

(ii) the current evidence for the existence of truly macroscopic quantum states is slim. Superpositions involving large numbers of Cooper pairs in a superconductor have been seen \cite{chiorescu03,ajl08}, although how macroscopic they are is still controversial \cite{whaley12,ajl16}. However, Cooper pair superpositions involve no mass displacements; and so far \cite{arndt} there is no evidence whatsoever for any position state superpositions involving masses larger than $\sim 10^5$ atomic units (ie., $\sim 10^{-22}~$kg, or $\sim 10^{-14}M_P$, where $M_P$ is the Planck mass).

(iii) Non-linearity in the Schrodinger equation is known to lead to inconsistencies with Bell inequalities and causality\cite{polchinski91,gisin90,scully90}. All Bell inequality experiments done so far on microscopic systems have verified the QM predictions; and no violation of causality has ever been found in physics.

We can divide theoretical attempts to modify QM by gravitational effects into 2 classes, according to whether they deal with non-relativistic QM or relativistic quantum fields.

The first of these starts with the non-relativistic Schrodinger equation; gravity is introduced using Newton's interaction potential. One set of such theories - the `collapse' theories - then introduces an {\it ad hoc} noise field, connected with gravity, which leads to wave-function collapse on a timescale depending on the mass of the object concerned \cite{diosi}. A related analysis of Penrose argues that the `mismatch' between, eg., 2 different branches of a quantum superposition can be related to an uncertainty in the proper time in these branches.

Although the Penrose and collapse models are physically quite different (there is no noise field in Penrose's discussion), they lead to a similar `Schrodinger-Newton' equation for the system dynamics - this non-linear equation is supposed to replace the Schrodinger equation. Experiments designed to test these models \cite{bouwm03} suffer from the fact that predictions vary widely depending on arbitrary assumptions
about, eg., how the mass distribution of solid bodies should be modeled
\cite{kleckner08}).

Clearly one would like a theory here which makes unambiguous predictions, and which is in some sense `natural', ie., it fits in naturally with those parts of physics that are already well established - including large parts of relativistic quantum field theory (QFT), as well as of classical GR. In our opinion the first attempt at such a theory was the remarkable early work of Kibble et al. \cite{kibbleb,kibble1,kibble2}, who attempted to bring in gravitation as the source of a non-linearity in relativistic QFT. As Kibble himself pointed out,  one inevitably finds inconsistencies between such non-linearity and the usual QM toolbox of measurements, operators, and Hilbert space. Kibble's work also stressed the inconsistencies involved in all ``semi-classical" treatments of quantum gravity, in which the non-linear dynamics is sourced by the expectation value $\langle T_{\mu\nu}(x) \rangle$ of the stress-energy tensor.

The work of Kibble - which has strongly influenced us - makes it clear that one not only requires some sort of `natural' mechanism for the breakdown of QM, but also a consistent theory - consistent not only internally, but with other parts of physics that are well established experimentally. It also makes it clear that if one is to go beyond QM, one will need a formal approach in which measurements, operators, etc., no longer play the central role that they do in standard QM.

\begin{figure}
\includegraphics[width=3.2in]{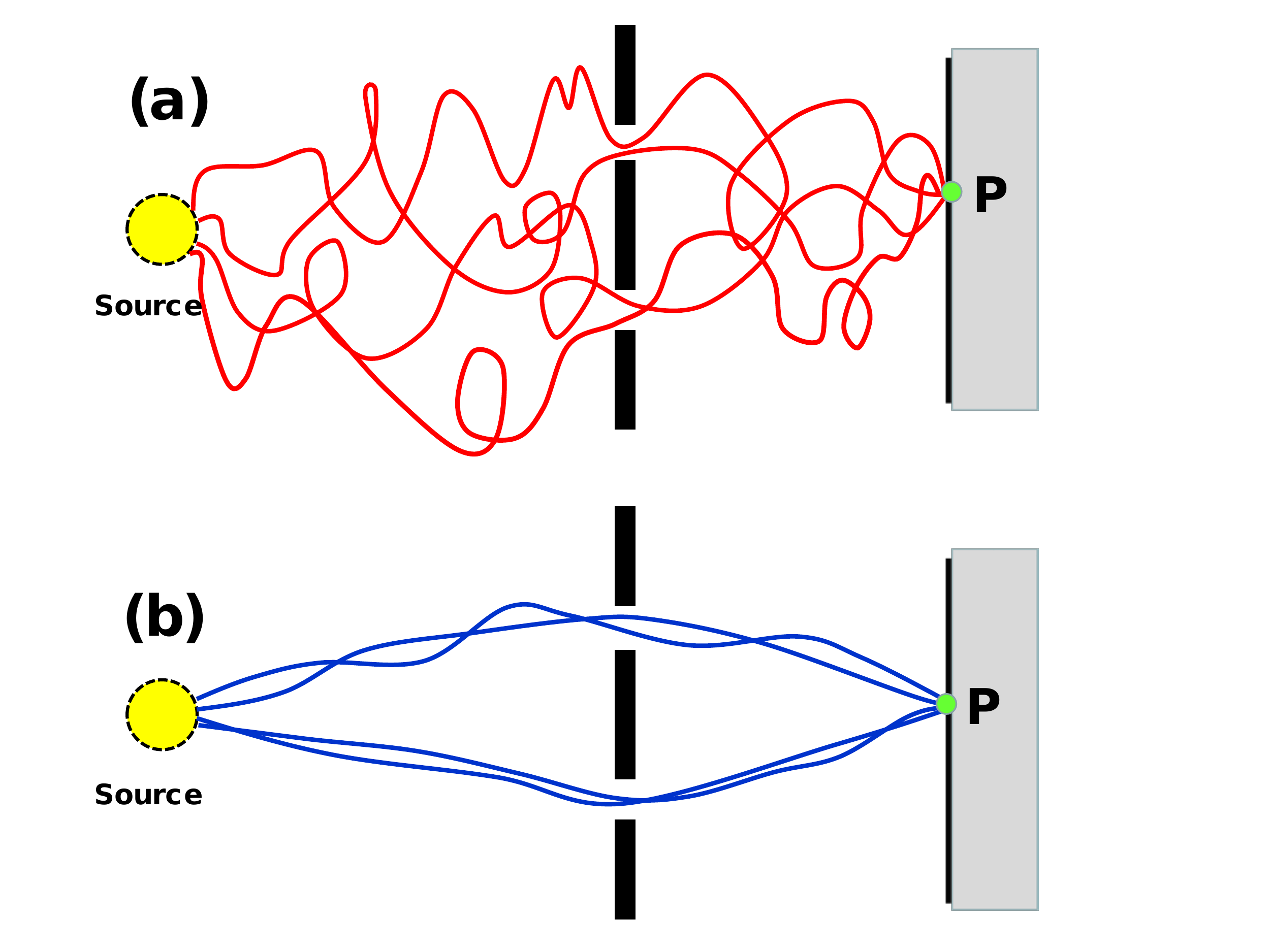}
\caption{\label{fig:2slit} The effect of interpath CWL gravitational correlations. In (a) some paths are shown for a QM particle moving through a 2-slit system from the source to a point $P$ on the screen. In (b) we have the same arrangement but with CWL correlations between the paths.  }
\end{figure}

The Correlated WorldLine (CWL) theory discussed in this paper is an attempt to find a consistent theory of quantum gravity in which gravity causes a breakdown on QM. CWL theory is formulated in terms of path integrals over both matter field and metric field configurations \cite{stamp12,stamp15,BCS18}. In CWL theory, QM breaks down because there are gravitationally-mediated correlations between all paths
in the path integral. These correlations then inevitably violate the superposition
principle, and lead to a ``path-bunching" effect \cite{stamp12,stamp15}, in which nearby paths in a path integral are attracted to each other (see Fig. \ref{fig:2slit}); this suppresses their usual tendency to spread over whatever domain may be accessible to them.


The `naturalness' of the theory comes because the form of the coupling between all paths is determined solely by the equivalence principle \cite{stamp15}. The CWL theory thus has no adjustable parameters - within the limitation that it is a low-energy theory (valid for energies $\ll M_P$, the Planck energy), it is self-contained, and requires no additional {\it ad hoc} ingredients. We re-emphasize the key role of the
path integral framework in the formulation of CWL theory \cite{stamp15}. Path integrals allow a more general formulation of QM than the usual wave function/state
formulation \cite{morette72}; and they can even go beyond QM, allowing us to formulate
theories in which paths are correlated, so that the superposition principle is violated.

In a previous paper \cite{BCS18} it was shown that there is a unique form of CWL theory (the `product' form) which has a sensible classical limit, and which appears to have a well-defined perturbation expansion in powers of the gravitational coupling (ie., in powers of the Planck length $l_P$). In the present paper we wish to explore the structure of this CWL theory, and establish a number of key conclusions about it. In particular, we want to derive the form of the semiclassical expansion about the classical limit; and we also want to show how one derives formal expansions in powers of $l_P$ about some classical background field configuration. Finally, we wish to show that the theory is gauge invariant (including invariance under diffeomorphisms), and satisfies all the Ward identities which express the conservation laws of the system.

These are all essential steps in (a) developing CWL theory as a practical tool for doing calculations of experimental phenomena, and (b) showing that it is internally consistent. We note that it is in general not a simple job to establish complete consistency for any field theory, and we cannot cover everything in this paper. In particular, one needs to define a consistent notion of causality; and we
would also like to understand if CWL theory is renormalizable. These questions are being dealt with separately.

In key parts of this paper we will use the formal covariant approach first introduced by DeWitt \cite{dewitt64,dewitt67b}, which allows very general investigations of, eg., gauge invariance, without having to specify objects like the action, or a state space.  As we shall see, this approach lends itself very naturally to theories going beyond standard QFT, and we will see how to extend it to CWL theory. For those unused to this approach, some explanation is given at various points in the text.

\subsection{Organization of Paper}
 \label{sec:org}

In the next section (section 2) we describe the basic structure of CWL theory - this is done by first comparing the generating functional $\mathbb{Q}[J]$ of CWL theory with the analogous functional ${\cal Z}[J]$ of conventional quantum gravity (where $J(x)$ is an external current coupling to the matter field). Both theories are defined here by path integrals; and since both require a proper treatment of boundary conditions and of gauge invariance, we then introduce the relevant terms in the action - including a third ghost field - as well as introducing our notation.

In section 3 we discuss semiclassical expansions about the classical limit of Einstein's theory. This is done as far as $\sim O(l_P^2)$, and displays the form of the relevant correlation functions involved, between matter, metric, and ghost fields.

In section 4 we show how one makes perturbative expansions in the gravitational coupling between matter and metric fields. This is done for the generating functional about a saddle point metric $g_0^{\mu\nu}(x)$, in powers of the Newton coupling $G$ (or equivalently, in terms of $l_P^2 = 16 \pi G \hbar$, where $l_P$ is the Planck length). Because we are dealing with three different fields, there is a proliferation of terms, even at order $l_P^2$; however this lowest order also yields the first term corresponding to a deviation from conventional quantum gravity, in the form of a path bunching term in the CWL generating functional. We give a detailed analysis of this term, and also show how it affects the correlation functions of the system (these being defined in the usual way as functional derivatives of the generating functional).

In section 5 we proceed to the most technically demanding part of the paper, the analysis of gauge dependence of the two gauge fields in the system (the metric field and the ghost field). We focus in this section on demonstrating that the CWL path bunching term is indeed invariant on-shell under a change of gauge conditions  -- this being the cornerstone of the Faddeev-Popov gauge fixing procedure.

One thing that we do not do in this paper is give explicit expressions for the field
propagators of the theory (these being the generalization to CWL theory of the Feynman
propagators for fields in conventional QFT). Such propagators are quite distinct from the field correlators (which were already discussed in our previous paper \cite{BCS18}); they require a rather lengthy treatment on their own, and are discussed in a separate paper \cite{jordan19}.


\section{Basic Structure of CWL Theory}
 \label{sec:CWLbasic}


In this section we review the basic structure of CWL theory in its product form
\cite{BCS18}. The detailed rationale for CWL theory has already been given in previous
papers \cite{stamp12,stamp15,BCS18}; and various arguments have also been given in the
past for some sort of theory of this kind \cite{kibbleb,penrose96,carney19}.

We begin with the form of the generating functional for CWL theory, which is used to
generate all correlation functions in a way similar to conventional quantum field theory (QFT). We specify the form of the action and boundary conditions on the fields, including ghost fields. Amongst other things, this allows us to compare CWL theory with conventional QFT, and establishes our notation.

\subsection{Generating Functional}
 \label{sec:CWLgenF}

One can write a generating functional for CWL theory in a way which parallels that in
conventional QFT.  We will formulate both theories in path integral theory language. Although the use of path integrals in discussing gravity can pose serious problems \cite{carlip01}, path
integrals nevertheless allow very general formulations of quantum field theories, and analysis of their consistency \cite{weinberg,dewitt03}; and they are also the natural language for
CWL theory.

\subsubsection{Conventional Theory}
 \label{sec:QFTgenF}

To define the conventional QFT of quantum gravity, let us consider a scalar field
$\phi(x)$ coupled to gravity, which in general has some self-coupling terms as well. Then one writes a generating functional in QFT of form
\begin{equation}
{\cal Z}[J] \;=\; \int Dg \, e^{iS_G[g]/\hbar} \; Z_M[\,g, J\,]
 \label{Zconv}
\end{equation}
in which $S_G[g] \equiv S_G[g_{\mu\nu}(x)]$ is the gravitational Einstein action, and the functional integration over the metric field is for the moment heuristic - we discuss it properly in the next sub-section. The quantity $Z_M[\,g, J\,]$, given by
\begin{equation}
Z_M[\,g, J\,]=\int D\phi\;e^{i(S_M[\,g,\phi\,]\;+\;\int J\phi)/\hbar}
 \label{Z-gj}
\end{equation}
is the generating functional for $\phi(x)$ in the presence of a {\it fixed}
$g_{\mu\nu}(x)$ and a fixed external source $J(x)$ coupling to $\phi(x)$. 

For most of this paper we will be looking at the limit  $J(x) \rightarrow 0$, where we can write
\begin{equation}
Z_M[g] \;=\; e^{i\,W_M[\,g\,]/\hbar} \;=\; \int D\phi \; e^{i\,S_M[\,\phi,g\,]/\hbar}
 \label{singlephi}
\end{equation}
so that
\begin{equation}
W_M[g] = -i \hbar \ln Z_M[g]
 \label{WM-g}
\end{equation}
is the generating functional for connected diagrams for the matter field in the presence of a frozen background metric field. In
the same way we define ${\cal W} = -i\hbar \ln {\cal Z}[J=0]$ as the total connected
generating functional, and $W_g = -i\hbar \ln Z_g$ as the connected generating functional for the gravitational field alone, with $Z_g = \int Dg \, e^{iS_G[g]/\hbar}$.

It is well known that the theory as written in (\ref{Zconv}) and (\ref{Z-gj}) is not
renormalizable, and that the path integrals suffer from various pathologies
\cite{carlip01}. However one can regard the path integrals as describing a low-energy
effective theory, where parameters like the gravitational coupling $G$ in the effective action are derived from experiment, and result from unknown very high-energy physics. In this case there is an effective high-energy cutoff built into the path integrals, and one can treat loops in a consistent and finite way \cite{donoghue15}. One also expects any very large changes in the metric to be eliminated (including those involving topology changes), so that problems defining the measure of the path integral are less severe. In what follows we will assume that we can implement a consistent quantization procedure under these circumstances.

\subsubsection{CWL Theory}
 \label{sec:CWLgenF}

Let us now consider the generating functional in CWL theory.In product CWL theory, 
because we are dealing with interactions between multiple paths
for the same particle and/or matter field, and hence with multiple copies of the matter
field, one starts by replacing the single scalar field $\phi(x)$ appearing in
(\ref{Zconv}) by a ``tower" $\Phi_n(x) \equiv \{ \phi_k(x) \}$ of multiple copies of
$\phi(x)$, with $k=1,2,...n$; and one then writes a generating functional
 \begin{eqnarray}
    &&\mathbb{Q}[J]=\prod\limits_{n=1}^\infty {\cal Q}_n[\,J\,], \nonumber\\
    &&{\cal Q}_n[\,J\,]=
    \!\int\! Dg\,e^{inS_G[\,g\,]/\hbar} \, \Big(Z_M\Big[\,g,
    \frac{J}{c_n}\Big]\Big)^n
 \label{bbQ-J1'}
    \end{eqnarray}
in which we take the product over all $n$, ie., we take the product of {\it all} the
towers $\Phi_n(x)$ of different $n$. In this formula $c_n$ is a regulating factor, which
increases with $n$ (and whose detailed behaviour we will uncover); and the gravitational
action in the $n$-th tower is rescaled by a factor $n$. This rescaling of $S_G[g]$ to
$nS_G[g]$ is equivalent to a scaling $G \rightarrow G/n$ for the metric $g_n$ in the
$n$-th tower, which reduces the effect of metric fluctuations at high $n$.

As discussed in \cite{BCS18} we can understand (\ref{bbQ-J1'}) a little better by also introducing a set of metric
fields $g_n$, with one such field for each tower of matter fields $\phi_k^{(n)}$, where
again $k=1,2,...n$. We can then write (\ref{bbQ-J1'}) in the form
\begin{equation}
\mathbb{Q}[J] \;=\;
\prod\limits_{n=1}^\infty
\int Dg_n \,e^{inS_G[\,g_n\,]/\hbar}
\left(Z_M\Big[\,g_n,
\frac{J}{c_n}\,\Big]\right)^n  \qquad
        \label{bbQ-J}
\end{equation}

Because the generating functional in (\ref{bbQ-J}) factorizes, we see that its logarithm
-- the generator of connected graphs -- is just a sum over single $g$ integrals, as in
(\ref{bbQ-J1'}), and we do not have correlations between $g_n$ and $g_m$ unless $n=m$. It
is then often easier to think of the generating functional in the form in
(\ref{bbQ-J1'}).

 We then have
\begin{equation}
    {\cal Q}_n[J] \;=\;
    \int Dg\, e^{in\left(S_G[\,g\,] + W_M[\,g,J/c_n\,]\right)/\hbar}  \label{Qn-W0}
\end{equation}

All of the theory in this paper will start from the generating functional $\mathbb{Q}[J]$ written as (\ref{bbQ-J1'}), with ${\cal Q}_n$ given by (\ref{Qn-W0}), or one of its equivalent forms. The correlation functions for this theory are given as functional derivatives of $\mathbb{Q}[J]$ with coefficients such that the correspondence principle is obeyed in the limit of vanishing gravitational interaction (see eqs.(\ref{1000})-(\ref{1000a}) below) -- this was described in detail in our previous paper \cite{BCS18}. One can also do semiclassical expansions (in $\hbar$) and perturbative expansions (in $l_P^2$) of the generating functional, by fairly simple adjustments to the usual techniques, which take into account the dependence of the action terms on $n$ in ${\cal Q}_n$.

These expansions reveal terms which have no analogue in any orthodox quantum field
theory; they come from the correlation between different paths in the path integral,
mediated by gravity. Following previous work \cite{stamp15} we will refer to these either as worldline correlations, or as ``path bunching" terms (since they cause paths in a path integral to congregate). At any order in $l_P^2$ we can isolate the terms contributing to this path bunching.

\subsection{Effective Action}
 \label{sec:action}

The discussion immediately above was somewhat schematic, because in doing the path integrals we need to deal with gauge and diffeomorphism invariance of the relevant fields. To do this we will use the standard device \cite{srednicki,FP67} of introducing in eqs.(\ref{bbQ-J1'})-(\ref{Qn-W0}) a gauge-fixing procedure and the relevant contribution from Faddeev-Popov ghost fields. We will also need to specify boundary conditions on the fields. In what follows we do this, and establish our notational conventions.

We will henceforth set the velocity of light $c = 1$, define the Planck length $l_P$ such that $l_P^2 = 16\pi G \hbar$, and write the gravitational action $S_G[g]$ as $S_G[g] = \hbar I[g]/l_P^2$. We will also write $\hbar = 1$ except when it is necessary, as in, eg., the discussion of semiclassical expansions.

To do the functional integration over the metric field $g_{\mu\nu}(x)$ we introduce a gauge-fixing function $\chi^{\mu}(g(x))$ in the usual way. We suppose that under a diffeomorphism  $x^{\mu}\rightarrow\xi^{\mu}+\xi^\mu(x)$, so that $g_{\mu\nu}(x) \rightarrow
g^{\xi}_{\mu\nu}(x)$, the action is invariant; to get rid of the gauge redundancy in the path integral under these transformations we add to the gravitational action a gauge-breaking term $\chi^\mu$, quadratic in gauge functions, and introduce into the integrand the  functional determinant
\begin{equation}
\Delta[g]={\rm Det}\,\Xi^{\mu}_{\nu}(x,x'|g),
\end{equation}
where the Faddeev-Popov ghost operator \cite{FP67} is
\begin{equation}
\Xi^{\mu}_{\nu}(x,x'|g) \;=\;\left.
\frac{\delta\chi^{\mu}(g^{\xi}(x))}{\delta\xi^{\nu}(x')}
\right|_{\,\xi = 0}\, .                                  \label{FPghost}
\end{equation}

We then have the conventional generating functional (\ref{Zconv}) in the form \cite{dewitt67c,mandelstam68,fradkinV73}
\begin{align}
{\cal Z}[J] \;&=\; \int Dg \, e^{i (S_G[g] + \frac{1}{2}\chi^{\mu}c_{\mu\nu}\chi^{\nu}- i {\rm Tr} \ln \Xi) }
\; Z_M[\,g, J\,]
 \label{Zconv2}
\end{align}
where we write $\rm {Det}\,\Xi = e^{Tr \ln \Xi}$. 

\begin{figure}
\includegraphics[width=3.2in]{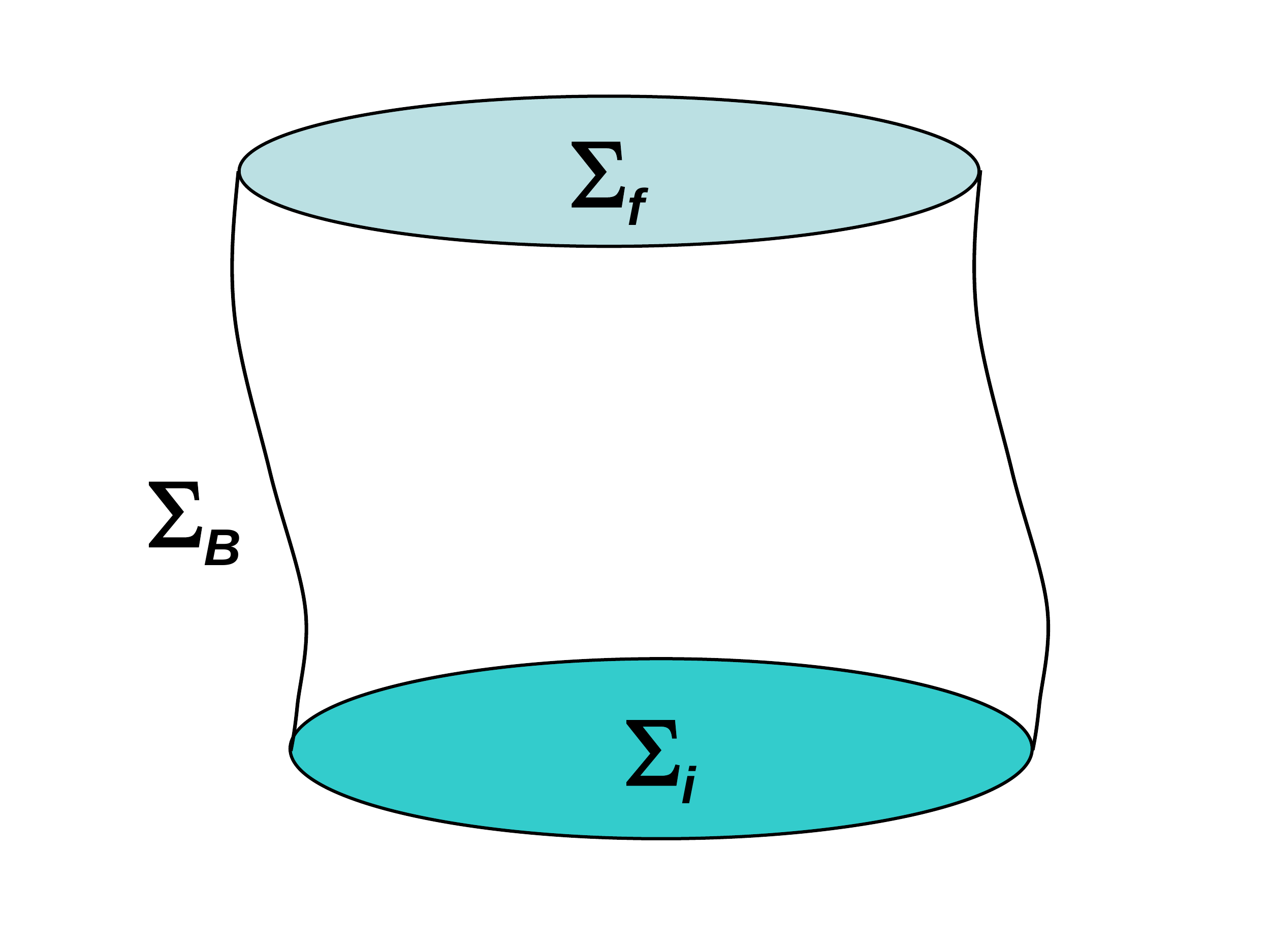}
\caption{\label{fig:Sigma} The boundary hypersurface $\Sigma$ of the spacetime region
considered in this paper. It comprises  future and past parts $\Sigma_f$ and $\Sigma_i$,
along with a region $\Sigma_B$ at spatial infinity. }
\end{figure}

To specify the form of the gravitational action we also need to specify the boundary
conditions for the system. The spacetime domain we will assume, in Fig. \ref{fig:Sigma}, involves a boundary hypersurface $\Sigma$ divided into future and past parts, along with a region $\Sigma_B$ at spatial infinity. Then, including the gauge-fixing term in the gravitational action, we write  $I[g] \equiv l_P^2 (S_G+\tfrac12\chi^{\mu}c_{\mu\nu}\chi^{\nu})/\hbar$ as
\begin{equation}
I[g] \;= \;  I_G^o + I_G^{YGH} + \tfrac{1}{2} \int \chi^{\mu}(x) c_{\mu\nu}(x,x')
\chi^{\nu}(x')
 \label{S-EH}
\end{equation}
where $I_G^o = \int g^{1/2}(x) R(x)$ is the bulk Einstein action, with $R(x)$ the Ricci scalar
curvature; where $I_{YGH}$ is the York-Gibbons-Hawking (YGH) boundary term \cite{YGH}; and where the
last term is the gauge-fixing term from (\ref{Zconv2}). The YGH term is required here
on the boundary hypersurface $\Sigma$, in order to cancel the 2nd derivative terms coming
from the bulk action; it is usually written as
\begin{equation}
I_{YGH} \;=\; 2\epsilon\, M_{P}^{2}\oint_{\Sigma}d^{3}y \sqrt{|\mathfrak{h}|}\,K
 \label{YGH-A}
\end{equation}
where $\mathfrak{h}$ is the determinant of the induced metric on $\Sigma$ (not to be
confused with $h(x)$, to be defined later as a gravitational fluctuation amplitude),
where $K$ is the trace of the extrinsic curvature $K_{ab}$ of $\Sigma$, and where
$\epsilon=\pm1$, depending on whether $\Sigma$ is timelike or spacelike. In the present paper, as well as in a subsequent paper on propagators in CWL theory \cite{CWL3}, we will be interested in initial states defined on an initial time slice,
which evolve to final states defined on a later time slice; and will assume that all fields vanish fast enough at $\Sigma_B$ that we can integrate by parts freely on spatial derivatives without picking up surface terms.

Clearly other spacetime boundary conditions could be assumed, leading to a different form for the effective action. Our choice here is largely determined by the ways in which we will be applying the theory in the near future, to problems involving experimental masses moving at velocities $\ll c$, where we are interested in calculating correlators, propagators, and other relevant quantities between 2 time slices. Note that we need the YGH term for such calculations even in the limit of linearized gravity.


Consider now the way in which gauge invariance is implemented. Under an infinitesimal diffeomorphism $x^{\mu}\rightarrow x^{\mu}+\xi^\mu(x)$, the
metric transforms as $g_{\mu\nu}(x)\rightarrow g_{\mu\nu}(x)+\delta_{\xi}g_{\mu\nu}(x)$
where
\begin{equation}
\delta_{\xi}g_{\mu\nu}
\;=\;\left(g_{\alpha\nu}\nabla_{\mu}
+g_{\alpha\mu}\nabla_{\nu}\right)\xi^{\alpha}.
 \label{d-xi-g}
\end{equation}
and we write this as
\begin{equation}
\delta_{\xi}g_{\mu\nu}(x)=\int d^{4}x'\,R_{\mu\nu,\alpha}(x,x'|g) \,\xi^{\alpha}(x')
 \label{defR}
\end{equation}
so that $R_{\mu\nu,\alpha}(x,x'|g)$, the generator of this infinitesimal gauge
transformation, is given by
\begin{equation}
R_{\mu\nu,\alpha}(x,x'|g)=\big(g_{\alpha\nu}(x')\nabla_\mu
+g_{\alpha\mu}(x')\nabla_\nu\big)\delta(x,x')
 \label{exp-R}
\end{equation}
The diffeomorphism invariance of the gravitational action under infinitesimal
transformations, viz., the statement that $I[g]=I[g+\delta_{\xi}g]$, can then be written in the form of a Noether identity as
\begin{equation}
\int d^{4}x\,\frac{\delta I[g]}{\delta g_{\mu\nu}(x)}
R_{\mu\nu,\alpha}(x,x'|g)=0
 \label{infDiff}
\end{equation}

It is convenient to choose a gauge-fixing function which is linearized about a background
field gauge $g_0$. This is typical strategy in gauge field theory - for example, in
Yang-Mills theory one often fixes $\bar{D}_{\mu}A^{\mu}_{\alpha}=0$ where $\bar{D}_{\mu}$
is the gauge covariant derivative with respect to a background gauge field. We then have
\begin{align}
\chi^{\mu}(g(x))\;&=\; \int d^4x'{\delta \chi^{\mu}(x) \over \delta g_{\alpha\beta}(x')}
\big(g_{\alpha\beta}(x')-g^{0}_{\alpha\beta}(x')\big) \nonumber \\
&\equiv\; \int
d^{4}x'\,\chi^{\mu,\alpha\beta}(x,x')\big(g_{\alpha\beta}(x')
-g^{0}_{\alpha\beta}(x')\big)
 \label{chi-kappa}
\end{align}
and the Faddeev-Popov operator becomes
\begin{equation}
\Xi^{\mu}_{\nu}(x,x'|g)=\int d^{4}z\,\chi^{\mu,\alpha\beta}(x,z)\,
R_{\alpha\beta,\nu}(z,x'|g). \label{17}
\end{equation}

The gauge fixing functional matrices $\chi^{\mu,\alpha\beta}$ and $c_{\mu\nu}$ are rather generic, apart from the requirement that the Faddeev-Popov ghost operator $\Xi^\mu_\nu$ and the matrix $c_{\mu\nu}$ are
both invertible. We also require that the matrix $c_{\mu\nu}$ is ultralocal, $c_{\mu\nu}(x,x')\sim\delta(x,x')$, for otherwise an extra ghost contribution $\sim {\rm Det}\, c_{\mu\nu}$ would be needed, which would not be proportional to the power divergent terms $\delta(0)$ usually discarded within dimensional regularization. We will write the inverse of $\Xi^{\mu\nu}$ as
\begin{equation}
\Xi^\mu_\nu\,\mathfrak{G}^\nu_\lambda=\delta^\mu_\lambda.
\end{equation}
which defines the Green function $\mathfrak{G}^\nu_\lambda$.

The notation so far will be familiar from QFT (see, eg., refs. \cite{srednicki,
weinberg}). It is also common in the literature to express formulae of this kind in an
abbreviated ``DeWitt" notation \cite{weinberg,dewitt64,dewitt67b,kallosh74}, in which
spacetime coordinates are incorporated into indices which also package together all other
variables. In this notation, for example, eqns. (\ref{infDiff}) and (\ref{17}) are
written as
\begin{eqnarray}
&&\frac{\delta I}{\delta g^a}R^a_\mu=0,\\
&&\Xi^{\mu}_\nu =\frac{\delta\chi^\mu}{\delta g^a}R^a_\nu \equiv
\chi^\mu_a R^a_\nu,
\end{eqnarray}
and we shall use this notation at various points in the paper - it is particularly useful
in the discussion of gauge invariance. A table with details of this notation appears in
DeWitt \cite{dewitt67b}.

In the CWL theory, we replace the scalar field $\phi(x)$ in the above by the set $\{
\Phi_n(x) \}$ of towers. This does not affect the definition of the ghost field, since
the only alteration to the gravitational part of the action is that  $I[g] \rightarrow
nI[g]$ (including the gauge-breaking term). Thus, for the $n$-th tower contribution ${\cal Q}_n$ in (\ref{bbQ-J1'}) to the
total generating functional, we have
\begin{equation}
{\cal Q}_n \;=\; \int Dg \, e^{i (nI[g]/l_P^2 - i{\rm Tr} \ln \Xi) } \left(Z_M[\,g_n,
J\,]\right)^n
 \label{Qn-FP}
\end{equation}
and we then carry on as before.

To summarize: we define quantum gravity, both conventional quantum gravity theory and CWL theory, by path integrals. Conventional quantum gravity is defined by (\ref{Zconv2}), and CWL theory by (\ref{bbQ-J1'}) and (\ref{Qn-FP}). The key mathematical objects in the CWL theory are

(i) the gravitational and matter actions $S_G[g]$ and $S_M[\phi, g]$, the
generating functional  $\mathbb{Q}[J]$, along with the functionals
$Z_M$, and $Z_g$, plus all their associated connected generating functionals; and

(ii) the gauge-fixing function $\chi^{\mu}(g)$ and the associated ghost operator
$\Xi^{\mu}_{\nu}(x,x'|g)$, along with the generator $R_{\alpha\beta,\nu}(z,x'|g)$ of
gauge transformations. Various derivatives of these objects, like
$\chi^{\mu,\alpha\beta}(x,x')$ or $\mathfrak{G}^\nu_\lambda(x,x')$, also figure in
the structure of the theory.

We see that the formal structure of CWL theory incorporates many objects from conventional quantum gravity. Because the action is unchanged from the conventional theory, we need the usual ghost field and gauge-fixing function appear; and the symmetries are also unchanged. The radical change appears in the form of the generating functional $\mathbb{Q}[J]$, and at first glance we might expect this to completely modify things like ghost fields. The reason that it does not is because, as noted earlier, the CWL coupling between the different field configurations or `copies' is fixed by the equivalence principle to be the same as that occurring inside each copy. Even the introduction of the factor $n$ multiplying the gravitational action $S_G[g]$ in the $n$-th tower of copies does not affect this.


\section{Classical Limit, Loop Expansion, and Correlators}
 \label{sec:semiC}


One thing that any CWL theory must do is have the correct classical limit as $\hbar
\rightarrow 0$, which in this case means it must reduce to Einstein's theory. Here we
will also demand that it have a well-behaved semiclassical expansion about this limit.

In this section we set up this expansion, around a well-defined classical saddle point of
the path integral for CWL theory. We  derive explicitly the form of the tree level and
1-loop contributions, which makes clear the overall structure of the expansion.

Note in passing here that a loop expansion is not the same as an expansion in powers of $\hbar$ (despite what appears in most quantum field theory texts). In conventional quantum gravity it was found long ago that loops contribute to, eg., classical perihelion precession \cite{iwasaki71}, and indeed loops in conventional QED can also give classical contributions \cite{donoghue04}. More recently, systematic treatments of higher-order loop contributions to classical gravitation have appeared \cite{cheung19}.


\subsection{Classical Limit}
 \label{sec:classL}


Although, as noted in the introduction, we do not in general have to distinguish between the different metric fields $g_n$, it is nevertheless interesting to look at the classical limit by starting from $\mathbb{Q}[J]$ written in the form (\ref{bbQ-J}). Consider now the saddle-point equations resulting from functional differentiation of $\mathbb{Q}[J]$ with respect to the different $g_n$; these equations read
    \begin{eqnarray}
    &&n\,\frac{\delta S_G[\,g_n\,]}{\delta g_n}
    +\sum\limits_{k=1}^n
    \frac{\delta S_M[\,g_n,\phi_i^{(n)}\,]}{\delta g_n}=0 \nonumber \\
    &&\frac{\delta S_M[\,g_n,\phi_k^{(n)}\,]}{\delta\phi_k^{(n)}}-{J \over c_n}=0
\label{saddleP}
    \end{eqnarray}

Now in the calculation of any path integral, or in the time evolution of the system, we
need to impose the same boundary conditions for all the different copies $\phi_k^{(n)}$
of the matter field. This is true first of all for all the $\phi_k^{(n)}$ inside a given
tower, so that the solutions for these equations coincide for all $k$ in the $n$-th
tower, ie., $\phi_k^{(n)}=\phi^{(n)}$. However this then means that all the $n$ matter
stress tensors in the saddle point equations must also be the same.  The coefficient $n$
in (\ref{saddleP}) then cancels out, and the first field equation in (\ref{saddleP})
gives
    \begin{eqnarray}
    \frac{\delta S_G[\,g_n\,]}{\delta g_n}
    +\frac{\delta S_M[\,g_n,\phi^{(n)}\,]}{\delta g_n}=0,
    \end{eqnarray}
which has the form of Einstein's field equation, with source field $\phi^{(n})$. Thus,
inside any tower $n$ we get the usual Einstein equation, sourced by the stress tensor of
a {\em single} matter field $\phi^{(n)}$.

In the absence of the source field $J(x)$ in the saddle point equations, things simplify
further: all reference to the tower index $n$ disappears, so that  $g_n$ and
$\phi^{(n)}$ satisfy the same set of equations for all $n$. Thus we can write, at the
$J=0$ saddle point, that
    \begin{eqnarray}
    &&\frac{\delta S_G[\,g_0\,]}{\delta g_0}
    +\sum\limits_{k=1}^n
    \frac{\delta S_M[\,g_0,\phi_0\,]}{\delta g_0}=0 \nonumber \\
    &&\frac{\delta S_M[\,g_0,\phi_0\,]}{\delta\phi_0}=0  \label{saddleP-E}
    \end{eqnarray}
in which $\phi^{(n)}=\phi_0$, and $g_n=g_0$, the classical solutions. These solutions are
then the starting point for a semiclassical expansion.

As we shall see, a key role in this expansion is played by the factor $n$ multiplying the
gravitational action. This leaves quantum fluctuation effects of the matter fields
untouched at high $n$; but it rescales the gravitational coupling constant $G$ for the
$n$-th metric $g_n$, so that $G\to G/n$. 

This weakening of the gravitational coupling
then reduces quantum fluctuation effects in $g_n$ at large $n$ (since the couplings
$\hbar \rightarrow \hbar/n$ and $l_P^2 \rightarrow l_P^2/n$). The reduction of
$\hbar$ to $\hbar/n$ then helps in any semiclassical expansion, since graviton loop
corrections and vertices are suppressed with growing $n$.


\subsection{Tree and 1-loop contributions}
 \label{sec:tree}


From the argument just given one might guess that the diagram rules for this theory can
be derived by simply rescaling $\hbar$ in the vertices and propagators. However this is
not correct, as we shall now see by calculating the terms explicitly, up to 1-loop order
in a semiclassical expansion.

\subsubsection{Contributions to $W_n$}
 \label{sec:1-loop}

Let us begin by writing the Faddeev-Popov term as $\ln\,\Delta[\,g\,] = {\rm Tr} \,\ln\Xi[\,g\,]$. Then the contribution ${\cal Q}_n$, given in heuristic form in (\ref{Qn-W0}), becomes
    \begin{align}
    {\cal Q}_n &\;=\;
    \int Dg\, e^{\frac{in}{\hbar}(I[\,g\,]/l_P^2 + W_M[\,g,J\,])}\,
    e^{{\rm Tr}\ln \varXi[\,g\,]} \nonumber \\
    &\; \equiv \; e^{i W_n/\hbar}\,.
     \label{Qn-W}
    \end{align}
Note that $W_n$ here refers to the contributions of all the fields in the $n$-th tower, not just the matter field.

The tree-level contribution to $W_n= nW = -i \hbar \ln {\cal Q}_n$ then reads
    \begin{eqnarray}
    W_n^{\rm tree}=nI[\,g_0\,]+\sum\limits_{i=1}^n S_M[\,\phi^i_0,g_0\,]=n(I+S_M),
    \end{eqnarray}
where, as shown above, the classical actions are calculated at the stationary points of
both gravity and matter field replicas defined by eqs.(\ref{saddleP-E}). Since the source (included in the matter action) is the same for all $i$, the matter contribution of a single field is just multiplied by
$n$.

The one-loop contribution is trickier. To shorten the expressions, we now use the compact
notation described in the introduction. We write
\begin{align}
    {{I}}_{ab} &\equiv \frac{\delta^2 I}{\delta g^a\,\delta
    g^b}\bigg{|}_{g=g_0; \; \phi^i = \phi^i_0} ; \nonumber \\
    {{S}}_{ab} &\equiv \frac{\delta^2 S_M}{\delta g^a\,\delta
    g^b}\bigg{|}_{g=g_0; \; \phi^i = \phi^i_0}
\end{align}
for the 2nd functional derivatives with respect to $g$; and for the derivatives involving
the matter field, we have
\begin{align}
    {{S}}_{ia} & \equiv \frac{\delta^2 S_M}{\delta\phi^i\,\delta
    g^a}\bigg{|}_{g=g_0; \; \phi^i = \phi^i_0} \nonumber \\
    {{S}}_{ik} & \;=\;\delta_{ik}\,{{S}}_{\phi\phi'} \equiv
    \frac{\delta^2 S_M}{\delta \phi^i\,\delta\phi^k}\bigg{|}_{g=g_0; \; \phi^i =
    \phi^i_0}
\label{deW-1}
    \end{align}

The second order functional derivatives of these equations, with respect to both $g$ and $\phi^i$, are fixed to their saddle point configurations. Note that ${{S}}_{\phi\phi'}=\delta^2 S_M/\delta\phi\delta\phi'|_{\phi^i =
\phi^i_0}$ is then the same for all $\phi^i$, again because at the stationary point of the path integral, all replica fields coincide.

We also introduce Green functions $D^{ac}$ and $G^{ik}$, defined by
    \begin{eqnarray}
    (\,{{I}}_{ab}+{{S}}_{ab})\, D^{bc}&=&\delta^c_a, \label{29}\\
     {{S}}_{ik}\,G^{km}&=&\delta_i^m.
    \end{eqnarray}
so that $D^{bc}$ is the graviton Green function, and $G^{km}$ the matter field Green
function, defined on a combined background of matter and metric fields.

We then find, for the 1-loop contribution, that
\begin{widetext}

    \begin{eqnarray}
    \!W_n^{\rm 1-loop} &=&-i\hbar{\rm Tr}\ln\varXi
    \;+\; {i \hbar \over 2}{\rm Tr}\ln\left[
    \begin{array}{cl}
    \!n({{I}}_{ab}+{{S}}_{ab}) &\! {{S}}_{ak}\\
    \;{{S}}_{ib}\;  & \!{{S}}_{ik}
    \end{array}
    \right] \nonumber \\
&=& -i\hbar{\rm Tr}\ln\varXi
    \;+\;
    {i\hbar \over 2} {\rm Tr} \bigg[ \ln\,
    ({{I}}_{ab}+{{S}}_{ab})
    \;+\; \ln\left(\,\delta^a_b
    -\frac1n\,D^{ac}\,{{S}}_{ci}G^{ik}{{S}}_{kb}\right) \bigg]
    \end{eqnarray}
\end{widetext}
where we note that ${\rm Tr}\,\ln\,
[\,n\,({{I}}_{ab}+{{S}}_{ab})\,]={\rm Tr}\,\ln\,
({{I}}_{ab}+{{S}}_{ab})$ up to an irrelevant $\delta(0)$-type
constant.

Notice now that because
$S_{ik}=\delta_{ik}\,S_{\phi\phi'}$ is diagonal, we have
$G^{ik}=\delta^{ik}\,G^{\phi\phi'}$, and
$S_{\phi\phi'}\,G^{\phi'\phi''}=\delta^{\phi''}_\phi$, and so the $1/n$
factor in the second determinant above completely cancels out.

It then follows that
$W_n^{\rm 1-loop}=W^{\rm 1-loop}$,
which is just the one-loop contribution of a theory without any CWL correlations, with a single matter field (ie., it is the 1-loop term for a conventional theory in which
gravitons couple to this matter field).

\subsubsection{Correlators}
 \label{sec:corr}

Consider now the form of the correlators that one derives from the connected generating functional. As we have just seen, $W_n^{\rm 1-loop}=W^{\rm 1-loop}$, and so from eqtns. (\ref{bbQ-J1'}) and
(\ref{singlephi}) we have
    \begin{eqnarray}
    &&W_{\rm CWL}^{\rm 1-loop}[\,J\,]=\sum\limits_{n=1}^\infty W^{\rm
    1-loop}\big[\,\frac{J}{c_n}\,\big],
    \end{eqnarray}
where $c_n$ is the regulator introduced in (\ref{bbQ-J1'}), and we explicitly show the
source dependence in the sum.

This infinite sum may be divergent, but the correlators generated by it are finite. The
correlators of the scalar matter field are given by the functional derivatives of
$\mathbb{Q}[J]$, in the form \cite{BCS18}
    \begin{eqnarray}
    &&\langle\,\phi(x_1)...\phi(x_l)\,
    \rangle^{CWL}_{\rm c}=\frac{{\cal G}_l(\{ x_k \})}{\sum\limits_{n=1}^\infty\,
    n c^{-l}_n},
    \label{1000}\\
    &&{\cal G}_l(\{ x_k \})=\left(\frac{\hbar}i\right)^n  \left.{ \delta^l
    \ln\mathbb{Q}[J] \over \delta J(x_1) .. \delta J(x_l)} \,\right|_{\,J = 0}
     \label{1000a}
\end{eqnarray}
so that we have
    \begin{eqnarray}
    &&\big\langle\,\phi(x_1)...\phi(x_l)\,
    \big\rangle^{CWL}_{\rm 1-loop}\nonumber\\
    &&\qquad=\frac{\sum\limits_{n=1}^\infty
    c_n^{-l}}
    {\sum\limits_{n=1}^\infty\,
    n c_n^{-l}}\,
    \big\langle\,\phi(x_1)...\phi(x_l)\,
    \big\rangle_{\rm 1-loop},              \label{1000b}
    \end{eqnarray}
where $\langle\,\phi(x_1)...\phi(x_l)\,\rangle_{\rm 1-loop}$ is the one-loop correlation function in conventional QFT without correlated world lines. This is convergent because of the regularization factors.

One can continue this expansion to higher-loop CWL theory correlators; the principles are
the same, so we do not give the details here.


\section{Perturbative Expansion in $l_P^2$ around the Saddle Point}
 \label{sec:pert}


In this section we discuss how to make expansions in the gravitational coupling $G =
l_P^2/16 \pi \hbar$ of the CWL generating functional. This expansion will be done around a
configuration $g_0 \equiv g^0_{\mu\nu}(x)$ of the metric field which gives a saddle point
in the action for the system - this configuration is of course not necessarily flat
space.

We will only go as far as $l_P^2$ in the expansion, because (i) there are many higher
order terms, the details of which require a paper of their own; and (ii) in discussing
experimental tests of the CWL theory, the terms $\sim O(l_P^2)$ turn out to be very
important, since it is at this order that the first correction to conventional quantum
gravity is found. This correction term, which correlates different matter paths, is the
lowest-order ``path bunching" term \cite{stamp15,BCS18}; it causes attractive
correlations between different paths. In the limit of low velocities, the path bunching
term gives the first correction to conventional quantum mechanics.

In what follows we begin by carrying out the formal expansion in $l_P^2$ on the $n$-th tower contribution ${\cal Q}_n$ to the generating functional, and exhibit all terms up to $\sim O(\l_P^2)$. There is a profusion of terms; we are interested here in the terms involving the matter field, and there are four of these. We focus on the ``path bunching" term that we find in this expansion, and give several explicit
expressions for it. Finally, we see how the path-bunching term affects the correlation
functions for the system.

\subsection{General Form of the Expansion}
 \label{sec:pert-G}

In this section we will again let $\Phi_n$  denote the full collection of fields at the
level of the $n$-th tower, so that $\Phi_n=\phi_1,...\phi_n$. We develop perturbation
theory in $l_P^2$ -- the gravitational coupling constant -- while keeping the path
integration over the matter field exact. This means that we change the order of
functional integration, and under the formal integral over $\phi$ we perturbatively
integrate over $g$.

In this section, and the next one, we will formulate the $l_P^2$-expansion in Euclidean spacetime. This is done to simplify the rather complex equations - we wish to avoid excessive use of powers of the imaginary unit $i$ which is characteristic of quantum mechanics in physical spacetime with a Lorentzian signature. After Wick rotation to a Euclidean theory, this difficulty does not arise. The Euclidean form will be particularly helpful when proving the gauge independence of the on-shell CWL effective action (see next section), which is an important part of the consistency check on the whole formalism. The return back to Lorentzian signature basically reduces to the replacement of the Euclidean quantities by the Lorentzian spacetime ones, by writing $S_G\to -iS_G$, $S_M\to -iS_M$, $I\to-iI$, etc.

For most of this section we will be dealing with the the $n$-th tower contribution ${\cal Q}_n[J]$ to the generating functional (cf. eqtns. (\ref{Qn-W0}) and (\ref{Qn-FP})). In Euclidean QFT this now reads
    \begin{equation}
{\cal Q}_n =\int D\Phi_n
    \int Dg \; e^{-{1\over \hbar} \left(nI[\,g\,]/l_P^2-\hbar {\rm Tr}\ln
    \varXi[\,g\,]+S_M[\,\Phi_n,g\,]\right)}
\label{Qn-Phin}
    \end{equation}
in which we again emphasize the rescaling of the gravitational action by a factor $n$.

To see the structure of the perturbative expansion, we write the metric field in an
expansion about the saddle point as $g=g_0+h$, and organize the integrand in powers of
the quantum field $h$. The background field $g_0$ again denotes the saddle point of the
path integral over $g$, so that $g_o$ is a solution of the {\em vacuum} Einstein equation
$\delta I[g_0]/\delta g_0=0$. In the expansion in powers of $l_P^2$, the matter stress
tensor is treated perturbatively and so, unlike in (\ref{saddleP-E}), it does not contribute to the saddle point configuration.

Functional differentiation, with respect to $\hbar$, of the various quantities in the integrand of the path integral then involves higher order vertices of general form
    \begin{equation}
{\cal O}_{a_1...a_n} =\left.\frac{\delta^n {\cal O}[\,g\,]}{\delta g^{a_1}...\delta
g^{a_n}} \right|_{\;g=g_0}
    \end{equation}
for some quantity ${\cal O}$ (which could be $I[g]$, $S_M[g.\phi]$, or ${\rm Tr}\ln
\varXi$), in which all functional derivatives are taken at the background gravitational
field $g_0$, whereas the matter field takes a generic value $\phi$ (to be integrated over
in the path integral).

\subsubsection{Expansion to order $l_P^2$}
 \label{sec:pert-G2}

It is useful, when we come to do the functional integration over $g(x)$ in
(\ref{Qn-Phin}), to introduce a simple notation for the terms that are produced. Let us
write
\begin{widetext}
    \begin{eqnarray}
    &&{\cal O}_{(n)}\equiv\left.\frac1{n!}\,\frac{\delta^n{\cal O}}{\delta g^{a_1}...\delta
    g^{a_n}}\right|_{\;g=g_0} h^{a_1}...h^{a_n},\\
    &&\langle\,{\cal O}_{(2n)}\,\rangle_h\equiv \left(\,{\rm Det}\,
    I_{ab}\,\right)^{1/2}\int Dh\,\exp\left(-\frac{1}{2\,l_P^2}\,I_{ab}\,h^a
    h^b\right)\,{\cal O}_{(2n)}[\,h\,]\propto l_P^{2n},\\
    &&\langle\,1\,\rangle=1,\qquad\qquad \langle\,h^ah^b\,\rangle_h=l_P^2\,
    D^{ab},\qquad\qquad\langle\,h^ah^bh^ch^d\rangle \;=\;
    l_P^4\big(\,D^{ab}D^{cd}+D^{ac}D^{bd}+D^{ad}D^{bc}\big), \quad ...\,.
    \end{eqnarray}
so that the bracketed subscripts denote the orders of the Taylor expansion in the quantum
field $h^a$, and angular brackets with the subscript $h$ denote the Gaussian integration
over $h^a$. Now doing the Gaussian integrals in (\ref{Qn-Phin}), we find
\begin{eqnarray}
    {\cal Q}_n&=&\exp\left[-\frac{nI_{(0)}}{l_P^2}+{\rm Tr}\,\ln \varXi_{(0)}-\frac12\,{\rm
    Tr}\,\ln\,\frac{\delta^2I_{(0)}}{\delta g_0\delta g_0}\right]\nonumber\\
    &&\qquad\times\int D\Phi_n\;e^{-S_M[\,\Phi_n,g_0\,]}
    \left\langle\,1+\frac{l_P^2}n\left[\,\frac{{I}_{(3)}^2}{2\,l_P^6}
    -\frac{{I}_{(4)}}{l_P^4}+\frac{\big({\rm Tr}\ln
    {\varXi}\big)_{(2)}}{l_P^2}+\frac{\big({\rm Tr}\ln
    {\varXi}\big)_{(1)}^2}{2\,l_P^2}-\frac{{I}_{(3)}\big({\rm Tr}\ln
    {\varXi}\big)_{(1)}}{l_P^4}\,\right]
    \right.\nonumber\\
    &&\qquad\qquad\qquad\qquad\qquad\qquad\left.+\frac{l_P^2}n\left[\,
    \frac{{S}_{(1)}^2}{2\,l_P^2}-\frac{{S}_{(2)}}{l_P^2}
    -\frac{{S}_{(1)}\big({\rm Tr}\ln
    {\varXi}\big)_{(1)}}{l_P^2}+\frac{{I}_{(3)}{S}_{(1)}}{l_P^4}\,\right]
    \;+\; O\big(\,l_P^4\,\big)\right\rangle_h,
\label{1order}
\end{eqnarray}

\end{widetext}
for the $n$-th tower contribution ${\cal Q}_n$ to the generating functional. For brevity, we have omitted the subscript $M$ in $S_{(1)}\equiv (S_M)_{(1)}$ and $S_{(2)}\equiv (S_M)_{(2)}$. 

In these averages,  $D^{ab}$ is the graviton Green's function, ie., the inverse of the
operator $I_{ab}$, so that
    \begin{eqnarray}
    I_{ac}D^{cb}=\delta^b_a,\quad I_{ab}\equiv\frac{\delta I}{\delta g^a\delta
    g^b}\,\Big|_{\;g=g_0}.
    \end{eqnarray}
Note that this graviton Green's function differs from the one defined in the previous section by eq.(\ref{29}) -- this is an artifact of the $l_P^2$-expansion, which is different from the $\hbar$-expansion because in the leading order it begins from the vacuum gravitational background.

Consider now the different terms in (\ref{1order}). The terms outside the integration
over the matter field just describe the background field $g_0$. At order $l_P^2$, the
group of 5 terms in the first square bracket involve the graviton field $h$ and the
Faddeev-Popov ghost field $\varXi$, but do not contain the matter field - these are just conventional quantum gravitational terms. Finally, the last group of 4 terms in the second square bracket does involve the matter field.

To make it clearer what is going on, let us again write ${\cal Q}_n$ in the form (\ref{Qn-W}), but now in Euclidean version, so that ${\cal Q}_n = e^{-W_n/\hbar}$. We can then write
\begin{equation}\label{Wn-GM}
W_n \;=\; W_n^{(g)} + W_n^{(M)}
\end{equation}
where the first gravitational term $W_n^{(g)}$ is derived from the integration over graviton and ghost fields, and can be written up to $O(l_P^4)$ as
\begin{equation}\label{Wn-G3}
W_n^{(g)} \;=\;\frac{n}{l_P^2}\,W^{\rm g}_{\rm tree}+W^{\rm g}_{\rm 1-loop} +\frac{l_P^2}n\,W^{\rm g}_{\rm 2-loop} 
\end{equation}
in which the tree contribution $W^{\rm g}_{\rm tree} = I_0$, and the other 2 terms from the integration over the graviton and ghost fields in the first two lines of (\ref{1order}). We shall not further investigate these gravitational terms here. The 2nd matter contribution $W_n^{(M)}$ to $W_n$ in (\ref{Wn-GM}) will be written up to $O(l_P^4)$ as
\begin{equation}
W_n^{(M)} \;=\; n\,W_M + W_n^{(corr)}
 \label{WnM2}
\end{equation}
where the first term is just the `bare' matter contribution without any fluctuation corrections, and the second term integrates the last line of (\ref{1order}) over graviton fluctuations and over the matter field $\Phi_n$. 

Let us write this latter term as
\begin{equation}
W_n^{(corr)} \;=\; -{l_P^2 \over n} \int D\Phi_n\;e^{-S_M[\,\Phi_n,g_0\,]}  C^{(M)}_n
 \label{int-QnM}
\end{equation}
in which $ C^{(M)}_n$ is the sum of the set of 4 correlators in the matter term, ie.,
    \begin{equation}
    C^{(M)}_n \;=\;
    \Big\langle\,\frac{{S}_{(1)}^2}{2\,l_P^2}-\frac{{S}_{(2)}}{l_P^2}
    -\frac{{S}_{(1)}({\rm Tr}\ln {\varXi})_{(1)}}{l_P^2}
    +\frac{{I}_{(3)}{S}_{(1)}}{l_P^4}
    \Big\rangle_h,
\label{QnM}
    \end{equation}
corresponding to the 4 matter terms in the last line (the 2nd square bracket) of eqtn. (\ref{1order}). It will be clear that $W_M$ denotes the contribution of the {\em single} matter field (\ref{Z-gj}) in the presence of a fixed metric background $g=g_0$, and  $C^{(M)}_n$ averages over the $n$ different fields in the $n$-th tower.


\begin{figure}
\includegraphics[width=3.2in]{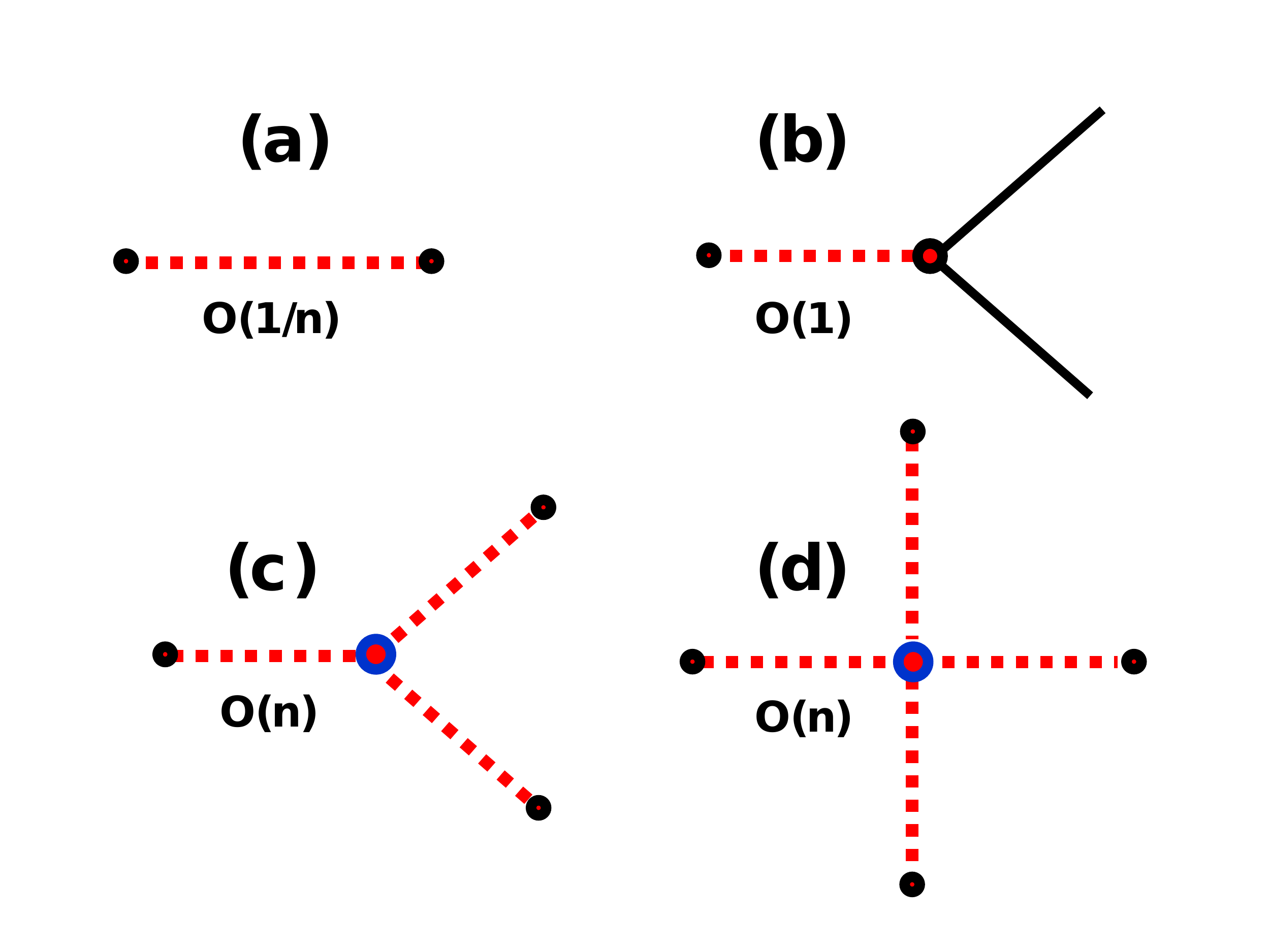}
\caption{\label{fig:CWL-vertex} Diagrams for some of the lower-order vertices in CWL theory, shown along with the order in $n$ that they carry when one sums over $n$ in expressions for the connected generating functional. In (a) we show the basic graviton propagator $D^{ab}$; diagrams (b), (c), and (d) show respectively a graviton-matter field interaction, a 3-point graviton self-interaction, and a 4-point graviton self-interaction. The smaller solid circle represents the graviton-matter interaction, the slightly larger circle represents graviton self-interactions, the solid line a matter field, and the hatched lines represent gravitons.    }
\end{figure}


\subsubsection{Diagrammatic Representation}
 \label{sec:pert-graph}

It is extremely useful to see how things are represented graphically, in dealing with these expressions. Before we do this, we emphasize that one must distinguish the diagram connectedness in pure matter theory on a fixed metric background from that of the full theory with quantum metric field. The functional (\ref{Wn-GM}) generates connected diagrams only if one includes and integrates over {\em all} propagators, including the graviton one', ie., if we take the logarithm of the full generating functional {\em after} integrating over the metric field, rather than before. Thus $W_n$ contains separate matter diagrams connected by the graviton lines, which decouple into disconnected pieces when breaking these graviton propagators.

To begin, we note that the form of the vertices in diagrammatic perturbation theory for CWL theory will look exactly the same as in conventional quantum gravity. This is of course because the action functional used in the two theories is the same. In Fig. \ref{fig:CWL-vertex} we show this for some of the vertices involving matter-graviton interactions, as well as for the bare graviton propagator. We could also show the vertices involving ghost fields, but we omit them in this figure.

However we also note that each line and vertex will depend on the tower index $n$. From either the original form of the action in, eg., eqtns. (\ref{Qn-W}) and (\ref{Qn-Phin}), we can determine the order in $n$ carried by any diagram when we do the final product over $n$ in the generating functional, or the sum over $n$ in the connected generating functional. 

We then see that since any free matter line represents $n$ copies, it automatically brings in a factor $n$ when summed over (so that the term $W_M$ in (\ref{WnM2}) is multiplied by $n$). From eqtn. (\ref{Wn-G3}) we also see that because graviton tree graphs carry a factor $n$, graviton vertices like those in Figs. \ref{fig:CWL-vertex}(c) and (d) must also carry a factor $n$. 

Unlike the matter field, neither gravitons nor ghosts are replicated. It then follows that the free graviton propagator in Fig. \ref{fig:CWL-vertex}(a) carries a factor $1/n$. This is also clear from the fact that adding graviton loops in $W_n^{(g)}$ in (\ref{Wn-G3}) lowers the order by a factor $1/n$ for each added loop.


\begin{figure}
\includegraphics[width=3.2in]{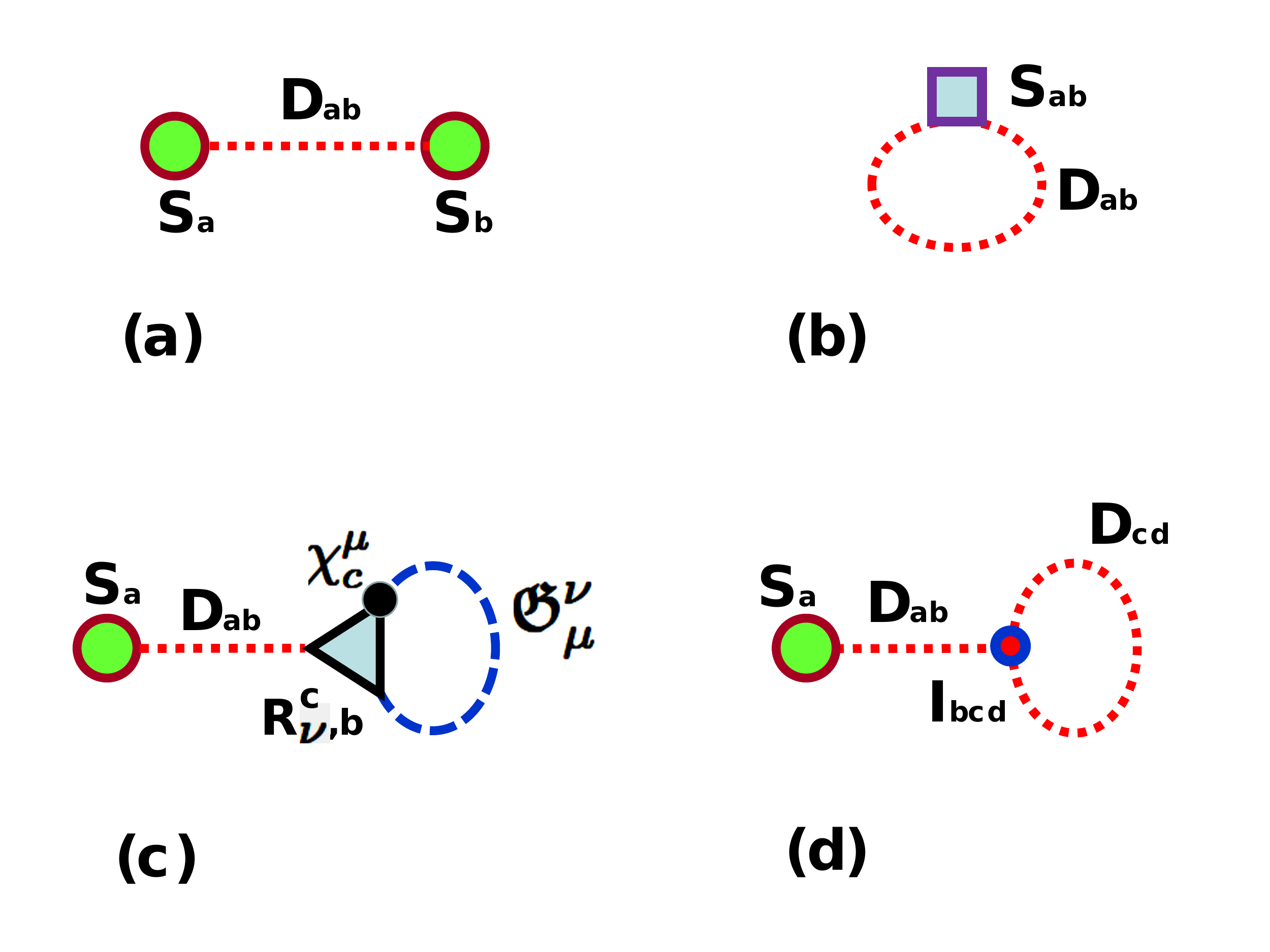}
\caption{\label{fig:Qdiagram} Diagrams for the matter terms ${\cal Q}_n^{(M)}$ in the
lowest order result for the $n$-th tower in $\mathbb{Q}$. In (a) the term correlating
matter paths, leading to ``path-bunching", is shown. The diagrams (b), (c), and (d) show correlations generated between the matter field and the graviton and ghost fields. The solid circles represent derivatives like ${S}_a$, the square the second derivative ${S}_{ab}$. Dotted lines show the graviton propagator $D_{ab}$, the dashed line the ghost propagator $\mathfrak{G}^{\nu}_{\mu}$; the small circle represents the graviton 3-point vertex ${I}_{abc}$, and the large triangle the 3-point vertex $R^c_{\nu,b}$. Finally, the solid dot represents the 2-point vertex $\chi^\mu_c$. For further explanation see text.  }
\end{figure}


With all this in mind, let us now return to the term $C^{(M)}_n$ that we found in eqtn. (\ref{QnM}). We can represent its contribution diagrammatically as shown in Fig. \ref{fig:Qdiagram}.  In this figure we see extra vertices over and above those in Fig. \ref{fig:CWL-vertex}), involving the ghost propagator and vertices between the ghost field and the matter and graviton fields.

The terms (a)-(d) in Fig.  \ref{fig:Qdiagram} are ordered following the terms in (\ref{QnM}). The most important of these 4 terms is the first one, quadratic in $S_{(1)}$. It describes a CWL correlation between 2 different worldlines of the $\phi$-fields, up to this order in $l_P^2$. Writing it explicitly, we have
\begin{equation}
{\cal E}_n \;=\; \Big\langle \frac{S_{(1)}^2}{2\,l_P^2}\,\Big\rangle_h
    \;\;=\;\; \frac12\,D^{ab} {S}_a{S}_b,
 \label{E-corr}
\end{equation}
where $S_a\equiv\delta S_M/\delta g^a$. As we discuss below in detail, it contains the lowest-order path-bunching effect.

The other terms in ${\cal Q}_n^{(M)}$ are linear in $S_{(1)}$ and $S_{(2)}$, ie., they
are linear superpositions of separate contributions of $\phi_i$ individually dressed by gravitons in the full set of fields $\Phi_n=\phi_1,\phi_2,...\phi_n$, but with no
graviton exchange between fields (so they do not correlate worldlines). We note from Fig. \ref{fig:Qdiagram} that whereas the first term (\ref{E-corr}) in $Q_n^{(M)}$ has a tree structure, the second term (\ref{II}) contains a graviton loop, while (\ref{III}) and (\ref{IV}) are tadpoles having the ghost and graviton loops respectively, with the attached graviton propagator carrying at its end the matter field object $S_a$.

Thus none of these other terms involves CWL correlations.
In terms of the graviton and ghost Green functions they read
    \begin{eqnarray}
    &&\Big\langle\,\frac{{S}_{(2)}}{l_P^2}\,
    \Big\rangle_h=\frac12\,{S}_{ab}D^{ab},        \label{II}\\
    &&\Big\langle\,
    \frac{{S}_{(1)}({\rm Tr}\ln {\varXi})_{(1)}}{l_P^2}\,
    \Big\rangle_h=
    {S}_aD^{ab}\chi^\mu_c
    R^c_{\nu,b}\mathfrak{G}^{\nu}_{\mu}   \;\;\;\;\;   \label{III}\\
    &&\Big\langle\,\frac{{I}_{(3)}{S}_{(1)}}{l_P^4}\,
    \Big\rangle_h
    =\frac12\,{S}_aD^{ab}{I}_{bcd}\,D^{cd},           \label{IV}
    \end{eqnarray}
where $R^c_{\nu,b}$ is the functional derivative of the gauge generator, ie.,
    \begin{equation}
    R^c_{\nu,b}\equiv\frac{\delta R^c_\nu}{\delta g^b}
 \label{DdiffG}
    \end{equation}
which is non-zero because the gauge algebra is non-Abelian. Note again that all three terms (\ref{II}), (\ref{III}), and (\ref{IV}) exist in conventional quantum gravity.

As just noted, the CWL term ${\cal E}_n$ in (\ref{E-corr}) is the one giving new physical effects. However, as we will see in section 5, even though we will not need to explicitly evaluate the other 3 terms, we do need to look at them when discussing the gauge invariance of the theory.


\subsection{Evaluation of lowest order terms}
\label{sec:pathB}


Now let us evaluate the 4 terms in the matter action just discussed, including  the CWL term ${\cal E}_n$ in (\ref{E-corr}). To evaluate these 4 terms we need to look at averages over the $n$ members of a given tower; in fact we need to evaluate terms of the form
\begin{equation}
   \big\langle\!\big\langle\, {\cal O}[\,\varPhi_n\,]\,\big\rangle\!\big\rangle\equiv
    \frac{\int D\varPhi_n\,e^{-S_M\big[\,\varPhi_n,g_0\big]}\,{\cal
    O}[\,\varPhi_n\,]}{\int D\varPhi_n\,e^{-S_M\big[\,\varPhi_n,g_0\big]}}
  \label{O-Phin}
\end{equation}
where the double angular brackets denote the quantum average of ${\cal O}[\,\varPhi_n\,]$ with respect to {\em all} quantum matter fields $\varPhi_n=\phi_1,...\phi_n$, $D\varPhi_n\equiv D\phi_1D\phi_2...D\phi_n$.

Now, since the multiple path integral factorizes as
    \begin{eqnarray}
    \int D\varPhi_n\,e^{-S_M\big[\,\varPhi_n,g_0\big]}  &=&\prod\limits_{i=1}^n
    \int D\phi_i\,e^{-S_M[\,\phi_i,g_0]} \nonumber \\ &=&Z_M^n[\,g_0]  \;=\;
    e^{-nW_M[\,g_0]}
  \label{fact1}
    \end{eqnarray}
the same must hold for quantum averages of products of observables with {\em different}
$\phi_i$, ie.,
    \begin{eqnarray}
    &&\big\langle\!\big\langle\, \prod\limits_{i}^n{\cal
    O}_i[\,\phi_i\,]\,\big\rangle\!\big\rangle=\prod\limits_{i}^n
    \big\langle\, {\cal O}_i[\,\phi\,]\,\big\rangle,
  \label{fact2}
    \end{eqnarray}
where $\big\langle\, {\cal O}[\,\phi\,]\,\big\rangle$ denotes the quantum average with
respect to a single matter field $\phi$, ie.,
    \begin{eqnarray}
    \big\langle\, {\cal O}[\,\phi\,]\,\big\rangle
    =\frac{\int D\phi\,
    e^{-S_M\big[\,\phi,g_0\big]}\,{\cal O}[\,\phi\,]}
    {\int D\phi\,
    e^{-S_M\big[\,\phi,g_0\big]}}.   \label{average_single}
    \end{eqnarray}

Let us now write $C^{(M)} \equiv  C^{(M)}_n[\,\varPhi_n\,]$ in terms of the stress-energy tensor. This is easy since each $S_a[\,\phi_i,g_0]$ is in fact the stress tensor (density) of the $i$-th matter field, ie. we have
    \begin{eqnarray}
    T^i_a=2\left.\frac{\delta S_M[\phi_i,g\,]}{\delta g^a}\right|_{\,g=g_0},
    \end{eqnarray}
and $\delta T^i_a/\delta g^b$ is the local ``seagull" vertex $\delta T^i_a/\delta
g^b=\delta T^i_b/\delta g^a \,=\, 2\delta^2 S_M[\phi_i,g\,]/\delta g^a\delta g^b$,

Bearing in mind that in (\ref{E-corr})-(\ref{IV}) we have
    \begin{eqnarray}
    S_a=\sum\limits_{i=1}^n S_a[\,\phi_i,g_0]
    \end{eqnarray}
it then follows that we can write
    \begin{eqnarray}
    \frac12\,S_a\,S_b\,D^{ab}=\frac1{8}\sum_{i=1}^n T^i_a\,T^i_b
    \, D^{ab}+\frac1{8}\sum_{i\neq j}^n T^i_a\,T^j_b\,D^{ab},
       \label{dblsum}
    \end{eqnarray}
where the diagonal $ii$ terms of the double sum (along with ``seagull" contributions)
represent of course gravitational dressing of separate matter world lines, whereas the
non-diagonal $i\neq j$ terms give the graviton entanglement of correlated world lines.
Then from (\ref{E-corr})-(\ref{II}) the matter term can be written in
terms of matter field quantum averages as
\begin{widetext}
    \begin{eqnarray}
    -\frac1n\,\big\langle\!\big\langle\,
    C^{(M)}_n \,\big\rangle\!\big\rangle &=&
    -\frac18\,n\,\big\langle\,T_a\,\big\rangle\,D^{ab}\,
    \big\langle\,T_b\,\big\rangle
    -\frac18\,D^{ab}\Big(\big\langle\,T_a\,T_b\,\big\rangle-
    \big\langle\,T_a\,\big\rangle
    \big\langle\,T_b\,\big\rangle\Big) \nonumber\\
    &&+\,\frac14\,\big\langle\,\delta T_a/\delta g^b\,\big\rangle\,D^{ab}
    +\frac12\,\big\langle\,T_a\,\big\rangle\, D^{ab}\chi^\mu_c
    R^c_{\nu,b} \mathfrak{G}^{\nu}_{\mu}
    -\frac14\,\big\langle\,T_a\,\big\rangle\, D^{ab}I_{bcd}\,D^{cd},
    \label{correlation}
    \end{eqnarray}
where the specific $n$-dependent coefficients come from the fact that the expectation
values $\big\langle\,T_a^i\,\big\rangle=\big\langle\,T_a\,\big\rangle$ and
$\big\langle\,T_a^i\, T_b^i\,\big\rangle=\big\langle\,T_a\,T_b\big\rangle$ coincide for different $i$.

Expanding out the notation, so that $D^{ab}\to
D_{\alpha\beta,\mu\nu}(x,y)$, and $T^i_a\to T^{\alpha\beta}_i(x)\; \equiv \; 2\,\delta
S[\phi_i,g\,]/\delta g_{\alpha\beta}(x)$, we then find that the first term in (\ref{correlation}) above - the path-bunching term - leads finally to a term $W_n^{CWL}$ in the correlated part $W_n^{(corr)}$ of the connected generating functional, given explicitly by
    \begin{eqnarray}
   W_n^{CWL} \;\;=\;\; -\frac{l_P^2}8\,n\,\big\langle\,T_a\,\big\rangle\,D^{ab}\,
    \big\langle\,T_b\,\big\rangle \;\;=\;\; -\frac{l_P^2}{8}\,n\,\int dx\,dy\,
    \big\langle\,T^{\alpha\beta}(x)\,\big\rangle\,D_{\alpha\beta,\mu\nu}(x,y)\,
    \big\langle\,T^{\mu\nu}(y)\,\big\rangle
    \label{main}
    \end{eqnarray}
\end{widetext}

We can in the same way expand the other terms in $\big\langle\!\big\langle\, C^{(M)}_n \,\big\rangle\!\big\rangle$, to find their contributions to $W_n^{(corr)}$.  However there is a key difference between the CWL-correlated term in (\ref{main}) and all the other terms in (\ref{correlation}), viz., the presence of the factor $n$ in (\ref{main}). This factor of $n$ comes from the double sum in (\ref{dblsum}), and is absent in the other terms in $W_n^{(corr)}$. As we will see, this is crucial in what follows.

The CWL path-bunching term ${\cal E}_n$  - is only the first correction to standard quantum gravity. At higher orders in $l_P^2$, such corrections proliferate; a proper enumeration of them all requires a lengthy analysis, which will be given elsewhere.

This concludes our analysis of the lowest order (in $l_P^2$) terms in a perturbative analysis of the generating functional for CWL theory.

\subsection{CWL Correlation Functions}
 \label{sec:corr}

Having dealt with the generating functional, we can now turn to the correlation functions that are derived from it by functional differentiation. In an earlier paper \cite{BCS18} the general form of these correlation functions was given already; see also eqtns. (\ref{1000})-(\ref{1000b}) above.

We now wish to see how the correlations functions are affected by the CWL term just computed. We therefore add the source term $-J\phi_i/c_n$ to the classical action of each field $\phi_i$, so that the average in (\ref{O-Phin}) becomes
    \begin{equation}
    \big\langle\!\big\langle\, {\cal O}[\,\varPhi_n\,]\,\big\rangle\!\big\rangle_J
    \equiv
    \frac{\int D\Phi_n\,e^{-S[\,\Phi_n\,]+J\sum_i\phi_i/c_n}\,{\cal
    O}[\,\Phi_n\,]}{\int D\Phi_n\,e^{-S[\,\Phi_n\,]+J\sum_i\phi_i/c_n}},
    \end{equation}
and a similar off-shell extension $\langle{\cal O}[\,\phi\,]\rangle\to\langle {\cal O}[\,\phi\,]\rangle_J$ holds for a single field average (\ref{average_single})). Therefore same factorization results as those in (\ref{fact1}) and (\ref{fact2}) apply.

Correlation functions are then given in the usual way by differentiating with respect to $J$, for which the simple rule
    \begin{eqnarray}
    \frac\delta{\delta J}\big\langle\, {\cal O}[\,\phi\,]\,\big\rangle_J=
    \frac1{c_n}\Big(\big\langle\, {\cal O}[\,\phi\,]\,\phi\,\big\rangle_J-
    \big\langle\, {\cal O}[\,\phi\,]\,\big\rangle_J\,
    \big\langle\,\phi\,\big\rangle_J\Big)        \label{61}
    \end{eqnarray}
can be used. We see here explicitly the subtraction of the disconnected part in any Feynman diagram for the correlators.

As an example, let us apply these results to the calculation of the two-field correlator up to $\sim O(l_P^2)$. By using the Euclidean version of (\ref{1000})-(\ref{1000a}) we have
\begin{widetext}
    \begin{eqnarray}
    \big\langle\,\phi_1\phi_2\,\big\rangle^{\rm CWL} \;\;&=&\;\;
    \frac1{\sum\limits_m
    mc_m^{-2}}\left.\sum\limits_{n=1}^\infty\frac{\delta^2
    W_n}{\delta J_1\,\delta J_2}
    \,\right|_{\,J=0}\nonumber\\
    &=& \;\; \big\langle\,\phi_1\phi_2\,\big\rangle
    \;-\; \frac{l_P^2}{\sum\limits_m
    mc_m^{-2}}\left.\sum\limits_{n=1}^\infty\frac1n\,\frac{\delta^2
    \big\langle\!\big\langle\, C^{(M)}_n[\,\varPhi\,]\,\big\rangle\!\big\rangle}{\delta
    J_1\,\delta J_2}
    \,\right|_{\,J=0}\, ,
    \label{phi-12}
    \end{eqnarray}
where the first three terms of $W_n$ defined by (\ref{Wn-GM}),(\ref{Wn-G3}) do not contribute at all (since they are independent of the source $J$), and the term $nW_M$  reproduces the correlator $\big\langle\,\phi_1\phi_2\,\big\rangle$ of a single quantum field in a fixed gravitational field -- the classical background $g=g_0$.

Using the expression (\ref{correlation}) and the rule (\ref{61}) we then get the answer as a sum of several terms which we present for a special case when they have a clear interpretation in terms of connected matter field graphs. This is the case of vanishing expectation value of $\phi$, $\big\langle\,\phi\,\big\rangle=0$, and vanishing full vertex $\big\langle\,T_{a}\,\phi\,\big\rangle=0$ (ie., corresponding to the case when the matter action $S_M[\phi,g]$ does not contain odd powers of $\phi$). One then finds
\begin{align}
\big\langle\,\phi_1\phi_2\,\big\rangle^{\rm CWL} \;=\;
    \big\langle\,\phi_1\phi_2\,\big\rangle
    & -\,\frac{l_P^2}4\,
    \big\langle\,T_a\,\phi_1\phi_2\,\big\rangle_{\rm c}
    D^{ab}\,\big\langle\,T_b\,\big\rangle \nonumber
    -\,\frac{l_P^2}8\,\frac{\sum\limits_n
    c_n^{-2}}{\sum\limits_n n c_n^{-2}}\,
    \Big(\,D^{ab}\,
    \big\langle\,T_a\,T_b\,\phi_1\phi_2\,\big\rangle_{\rm c}
    \\
    &-2D^{ab}\,
    \Big\langle\,\frac{\delta T_a}{\delta g^b}\,\phi_1\phi_2\,\Big\rangle_{\rm c}
    -4
    \big\langle\,T_a\,\phi_1\phi_2\,\big\rangle_{\rm c}D^{ab}\,
    \chi^\mu_c
    R^c_{\nu,b}\mathfrak{G}^{\nu}_{\mu}
    +2
    \big\langle\,T_a\,\phi_1\phi_2\,\big\rangle_{\rm c}
    D^{ab}\,I_{bcd}\,D^{cd}\,\Big),
 \label{CWL-12}
\end{align}
Here the subscript ``c" denotes the connected part of the relevant single field correlators, which in the aforementioned case reduces to the set of expressions
    \begin{eqnarray}
    &&\big\langle\,T_a\,\phi_1\phi_2\,\big\rangle_{\rm c}=
    \big\langle\,T_a\,\phi_1\phi_2\,\big\rangle
    -\big\langle\,T_a\,\big\rangle
    \big\langle\,\phi_1\phi_2\,\big\rangle,             \label{CWL-exp1}\\
    &&\big\langle\,T_a\,T_b\,\phi_1\phi_2\,\big\rangle_{\rm
    c}=\big\langle\,T_a\,T_b\,\phi_1\phi_2\,\big\rangle-
    \big\langle\,T_a\,T_b\,\big\rangle\,\big\langle\,\phi_1\phi_2\,\big\rangle
    -2\,\big\langle\,T_{(a}\,\big\rangle\,\big\langle\,T_{b)}\,
    \phi_1\phi_2\,\big\rangle
    +2\,\big\langle\,T_a\,\big\rangle\,\big\langle\,T_b\,\big\rangle\,
    \big\langle\,\phi_1\phi_2\,\big\rangle,              \label{CWL-exp2}\\
    &&\Big\langle\,\frac{\delta T_a}{\delta g^b}\,\phi_1\phi_2\,\Big\rangle_{\rm c}
    =\Big\langle\,\frac{\delta T_a}{\delta g^b}\,\phi_1\phi_2\,\Big\rangle-
    \Big\langle\,\frac{\delta T_a}{\delta g^b}\,\Big\rangle
    \big\langle\,\phi_1\phi_2\,\big\rangle.                \label{CWL-exp3}
    \end{eqnarray}

\end{widetext}


\begin{figure}
\includegraphics[width=3.2in]{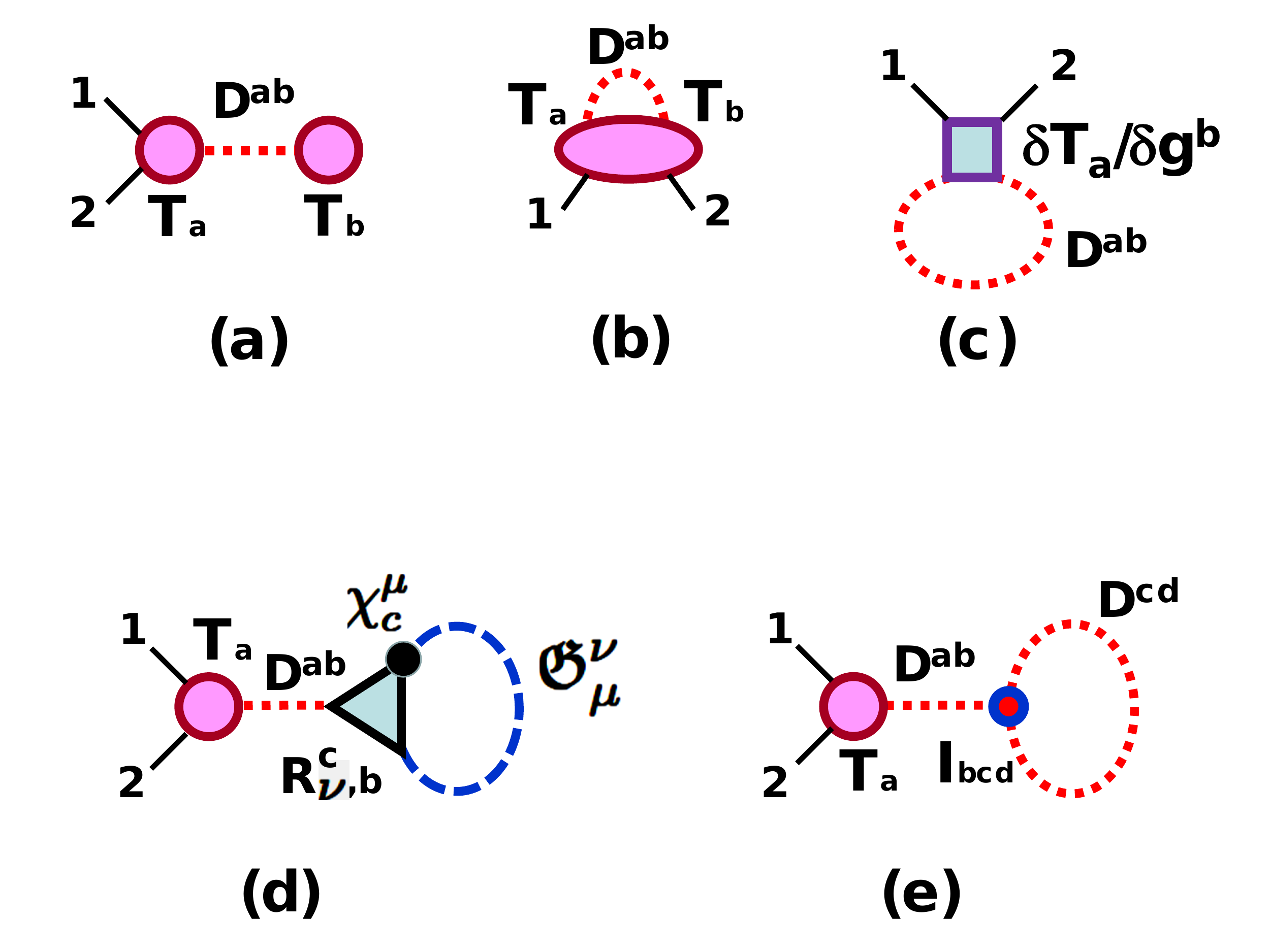}
\caption{\label{fig:QcorrelatorB} Diagrams for the different terms in the correlator $\big\langle\,\phi_1\phi_2\,\big\rangle^{\rm CWL}$ in eqtn. (\ref{CWL-12}), between field configurations $\phi_1(x)$ and $\phi_2(x)$ (labelled as ``$1$" and ``$2$" in the diagram, and shown as straight solid line insertions). All terms in eqtn. (\ref{CWL-12}) are shown in the order in which they appear, except for the first term $\big\langle\,\phi_1\phi_2\,\big\rangle$.  In (a) we have the ``path-bunching" term, while (c) shows the `seagull' contribution, and the diagrams in (d) and (e) show contributions coming from the coupling between the matter field and the graviton and ghost fields. The solid circles represent expectations of stress-energies like $T_a$, the elliptical solid the combined correlator of $T_a$ and $T_b$, and the square shows the derivative $\delta T_a/\delta g^b$. Dotted lines show the graviton propagator $D_{ab}$, the dashed line the ghost propagator
$\mathfrak{G}^{\nu}_{\mu}$; the small circle represents the graviton 3-point
vertex ${I}_{abc}$, and the large triangle the 3-point vertex $R^c_{\nu,b}$. Finally, the solid dot represents the 2-point vertex $\chi^\mu_c$. For further explanation see text.  }
\end{figure}


It is very useful here to represent the different terms diagrammatically - see Fig. \ref{fig:QcorrelatorB}. The diagram rules are the same as those in Fig. \ref{fig:Qdiagram}, except that now we add external insertions corresponding to the field configurations $\phi(x_1)$ and $\phi(x_2)$ (labeled as ``$1$" and ``$2$" in the diagram), whose mutual correlation we are asking for. If we now go through the different terms in (\ref{CWL-12}) we have the following contributions:

(i) the `free' correlator $ \big\langle\,\phi_1\phi_2\,\big\rangle$, ie., the correlator without any gravitational interactions (if the scalar field has, eg., a $\phi^4$ self-coupling in it, then this `free' correlator would also include these self-interactions). This graph is not shown in Fig. \ref{fig:QcorrelatorB} (it would simply appear as a black line connecting $\phi_1(x)$ and $\phi_2(x)$). This contribution is of course entirely conventional.

(ii) the path-bunching term of the tadpole structure, in Fig.\ref{fig:QcorrelatorB}(a) in which a path $b$ decorates, via the interaction $D^{ab}$, the path $a$ on which correlations are being determined ($T_a$ and $T_b$ belonging to different matter species associated with these paths). The diagram here represents this term in a compact connected form, implying the subtraction of disconnected parts as in eqtn. (\ref{CWL-exp1}). Note that in conventional quantum gravity there is also a contribution of this form; however $a$ and $b$ entries then belong to one and the same single matter field, so that one then has $a=b$. 

(iii) a `self-energy' graph without path-bunching -- both $T_a$ and $T_b$ belonging to one and the same matter replica -- shown in Fig.\ref{fig:QcorrelatorB}(b), which again implies the subtraction of disconnected parts as in eqtn. (\ref{CWL-exp2}).

(iv) the seagull graph, in Fig.\ref{fig:QcorrelatorB}(c), which involves the 4-vertex $\delta^2(\delta T_a/\delta g^b)/\delta\phi\,\delta\phi\equiv  2\delta^4S_M/\delta g^a\delta g^b\delta\phi\,\delta\phi$. This contribution is also familiar from conventional quantum gravity.

(v) the  2 'tadpole' diagrams in Figs.\ref{fig:QcorrelatorB}(d) and (e); these contain ghost and graviton loops. Again, terms of this form, with  $a$ and $b$ entries belonging to a single matter field, are familiar from conventional quantum gravity.

Returning now to eqtn. (\ref{CWL-12}), we observe that, as expected, all of these  diagrams except the path-bunching tadpole diagram in Fig. \ref{fig:QcorrelatorB}(a) are suppressed by the 'normalization' factor $\chi$ given by 
    \begin{eqnarray}
   \chi \;=\;  \frac{\sum\limits_{n=1}^\infty
    c_n^{-2}}{\sum\limits_{n=1}^\infty
    n c_n^{-2}} \;\; < \;\; 1.       \label{factor}
    \end{eqnarray}
Only the conventional term $\big\langle\,\phi_1\phi_2\,\big\rangle$ coming from $W_M$, and the path-bunching tadpole diagram coming from $W^{CWL}$ in Fig.\ref{fig:QcorrelatorB}(a) are not suppressed by $\chi$. This is in full accordance with our preceding paper \cite{BCS18}.

From the diagrammatic point of view, what distinguishes the CWL path-bunching term in Fig. \ref{fig:QcorrelatorB}(a) from the other graphs is that, as noted above, it does not take the usual 'tadpole' form, with the entries 1 and 2 belonging to the same replica as $\langle\,T_b\,\rangle$. The summation over $n$ replicas of $\langle\,T_b\,\rangle$ different from those of $\langle\,T_b\phi_1\phi_2\,\rangle$ then gives an extra coefficient of $n$, which explains the absence of suppression of this diagram by the factor $\chi$ in (\ref{factor}). This coefficient $n$ cancels the coefficient $1/n$ in the reduced gravitational coupling $l_P^2/n$. Note that for graviton and ghost tadpoles this mechanism does not work, because unlike the $n$ replicas in the matter contribution $\langle\,T_b\,\rangle$, graviton and ghost loop graphs contain only one un-replicated graviton and one set of un-replicated Faddeev-Popov ghosts.


\section{Gauge Dependence and Ward Identities}
 \label{sec:gauge}


The discussion of gravitational gauge invariance is notoriously difficult. Ordinary flat space QFT can deal with gauge invariance in various ways - by, eg., defining `physical states' \cite{mandelstam68,dirac55,mandelstam62,rossi80}, or, in path integral theory, by using a Faddeev-Popov procedure \cite{FP67}. However in quantum gravity things are more complicated - one would like to define meaningful local physical observables, but this is incompatible with diffeomorphism invariance. In spite of this, attempts to define physical states have been made \cite{mandelstam62,giddings}, and various ways of defining path integrals for quantum gravity have been given \cite{dewitt64,dewitt67b,FP-grav,teitelboim,bern,rovelli}.

The corner stone of these definitions is the requirement of on-shell gauge independence of the path integral, which guarantees the uniqueness of the resulting physical S-matrix. The general non-perturbative proof of this property for path integrals \cite{dewitt64,dewitt67b,FP-grav,teitelboim} equally well applies to its CWL version simply because each of the CWL factors ${\cal Q}_n$ in (\ref{Qn-W}) already takes the form of a conventional path integral for quantum gravity, in which a single metric field couples to $n$ matter field replicas with a standard Faddeev-Popov gauge fixing procedure. The mechanism of this gauge independence can then be checked order by order in perturbation theory, and the demonstration that the purely gravitational tree, one-loop and two-loop terms of in $W_n$ are gauge independent, when on shell \cite{barvinsky-vilkovisky87,dewitt03}, follows conventional lines.

However there is also the non-trivial CWL matter contribution $W^{CWL}$, which first manifests itself at order $l_P^2$, and which explicitly involves the gauge conditions $\chi^\mu$, and its gauge independence at $J=0$ is therefore far from being obvious. Since this contribution is tied to the CWL path-bunching effect, the question of its gauge dependence becomes very important.

Thus, in what follows our main goal is to show that in a path integral formulation, gauge independence in CWL theory can be formulated and proven in a way analogous to that in conventional quantum gravity. We stress again that we are dealing with a low-energy effective theory, and so we do not address questions surrounding the correct definition of local observables in CWL theory.

We begin, in sub-section A below, by recalling how gauge and diffeomorphism invariance are formulated for a path integral theory of conventional quantum gravity \cite{dewitt67b}. We then adapt this treatment to CWL theory, and then we show how in the lowest CWL correction to conventional quantum gravity, gauge invariance goes through as before. Although the demonstration is technically tedious, the basic idea is straightforward - essentially we want to see that the `relative phases' between two or more correlated paths in a CWL term do not mess up gauge invariance.

\subsection{Gauge dependence for metric and ghost field objects}

Let us first recall how, in conventional quantum gravity, one characterizes the gauge dependence of objects like Green functions, or contributions to the effective action, under a change of gauge conditions in the Faddeev-Popov gauge fixing procedure.  We will again, in order to streamline the discussion, use the condensed DeWitt notation already noted in section II.B.

The diffeomorphism invariance of the pure gravitational action is expressed by the Noether identity (\ref{infDiff}), which is written in condensed form as
    \begin{eqnarray}
    R^a_\mu S^G_a=0.
 \label{60}
    \end{eqnarray}

In the same way, gauge invariance for the matter action $S_M=S_M[\,\phi,g\,]$ involves gauge transformations of both gravitational and matter fields, so that
    \begin{eqnarray}
    R^a_\mu S_a+R^\phi_\mu S_\phi=0,
    \end{eqnarray}
where, as discussed in section II.B., $R^\phi_\mu$ denotes the generator of the gauge transformation of $\phi$,
and $S_\phi\equiv\delta S_M/\delta\phi$.

These identities hold for all field configurations, including off-shell ones. They can then be used to generate, eg., Ward identities for bare vertices, which follow from functionally
varying  (\ref{60}). We then get
    \begin{eqnarray}
    R^a_\mu S_{ab}^G &=&-R^a_{\nu,b}S_a^G \nonumber \\
    \,\, R^a_\mu S_{abc}^G &=& -R^a_{\nu,b}S_{ac}^G-R^a_{\nu,c}S_{ab}^G
    \end{eqnarray}
as well as a combined Ward identity for gauge and ghost propagators, viz.,
    \begin{equation}
c_{\mu\nu}\chi^\nu_a\,D^{ba}\;=\; \mathfrak{G}^{\nu}_{\mu}\,R^a_\nu + \mathfrak{G}^{\nu}_{\mu}\,R^b_{\nu,c}\,S_b^G\,D^{ca}\,
 \label{G+gh-WI}
    \end{equation}

For on-shell gravitational configurations, ie., those for which $g = g_0$, the last term in (\ref{G+gh-WI}) vanishes, and $S_b^G = 0$, and so we are thus led to an identity relating the gauge and ghost propagators taking the simple form:
    \begin{equation}
c_{\mu\nu}\chi^\nu_a \,D^{ba} =
    \mathfrak{G}^{\nu}_{\mu}\,R^a_\nu
 \label{W1}
    \end{equation}

In all discussion from now on we will calculate all the quantities on shell, that is for $g=g_0$. We then have the graviton
operator
    \begin{eqnarray}
    I_{ab}=S^G_{ab}+\chi^\mu_a\,c_{\mu\nu}\chi^\nu_b\,,
    \end{eqnarray}
and the variation of its Green function with respect to the infinitesimal change of the
gauge conditions matrix $\chi^\mu_a$ reads
    \begin{eqnarray}
    \delta_\chi D^{ab} &=&-2\,D^{(ac}\,\chi^\mu_c\,c_{\mu\nu}\,\delta\chi^\nu_d D^{db)}
    \nonumber \\ &=& -2\,R^{(a}_\nu\,\mathfrak{G}^{\nu}_{\mu}\,\delta\chi^\mu_d\,D^{db)}\,,
    \label{deltaG}
    \end{eqnarray}
where the round brackets around two indices imply symmetrization, so that, eg., $X^{(ab)}= \tfrac{1}{2} (X^{ab}+X^{ba})$.

All path integrals for the action are gauge independent on shell, ie., for $g=g_0$ and with sources switched off. The same should hold order by order in an $l_P^2$-expansion, ie., we expect that if we sum all diagrams contributing to a given order in $l_P^2$, this sum will be gauge invariant, even if individual diagrams are not.

\subsection{Gauge independence in CWL theory}

If we are to have gauge invariance of CWL theory, we also expect this to hold at any given order in an $l_P^2$ expansion of the generating functional. We now wish to investigate this. Clearly any proof should be independent of the number of paths that are correlated, ie., of the number $n$ of copies or `replicas', and so, as before, we denote all of them by one symbol $\Phi$.

In what follows we will look at the lowest non-trivial term in the $l_P^2$ expansion, ie., the matter term ${\cal Q}_n^{(M)}$ in the effective action in ${\cal Q}_n$. As we saw earlier, this term contains both a CWL path-bunching or `entanglement' term ${\cal E}$, plus three other conventional terms - see eqtns. (\ref{E-corr})-(\ref{IV}).

The proof of gauge independence of the CWL theory at order $l_P^2$  begins with the observation that the contribution ${\cal E}$ in (\ref{E-corr}) or $W^{CWL}$ in (\ref{main}) has the structure of the simplest tree-level $2 \to 2$ graviton scattering amplitude -- two stress tensors mediated by the graviton propagator. Since the change of this amplitude under the variation of the gauge conditions (\ref{deltaG}) is proportional to the diffeomorphism generator $R^{a}_\nu$, we might expect ${\cal E}$ and $W^{CWL}$ to be gauge independent in view of the stress tensor conservation identity
\begin{equation}
R^{a}_\nu S_a \;=\; -2\nabla^\mu T_{\mu\nu} \;=\; 0
 \label{stressC}
\end{equation}

However, this conservation law holds only on shell when the matter equations of motion are enforced; it does not hold before path integration over $\phi$ is carried out. Therefore, the mechanism of gauge independence is a little more subtle, and to do things properly we need to evaluate all the diagrams contributing to ${\cal Q}_n^{(M)}$, ie.,, all the contributions (\ref{E-corr})-(\ref{IV}). We now take these in turn.

We will find that gauge independence holds separately for the CWL contribution ${\cal E}_n$ in (\ref{E-corr}) and for the other 3 terms contributing to
${\cal Q}_n^{(M)}$. This is good, because as we have seen in the last section, these terms contribute differently to $W_n$.

\subsubsection{CWL path-bunching term}
 \label{sec:pathB-gauge}

We consider the first term of the  correlation (\ref{QnM}) appearing in the exponentiated matter action  (\ref{int-QnM}), which leads to the CWL path-bunching or ``entanglement" term (\ref{E-corr}). Let us take the integrand of the integral over the matter fields, and then vary it; this gives
    \begin{eqnarray}
    \delta_\chi\Big\langle\,\frac{S_{(1)}^2}{2\,l_P^2}\,\Big\rangle_h\,e^{-S_M}
    &=&\tfrac{1}{2}\,\delta_\chi (D^{ab}S_aS_b)\,e^{-S_M}
    \nonumber \\
    &=& e^{-S_M}\, (\mathfrak{G}^{\nu}_{\mu}\,D^{cb} S_b\,R^\phi_\nu S_\phi)\,\delta\chi^\mu_c \qquad
 \label{d-chiS2}
\end{eqnarray}
where we have used (\ref{deltaG}) for the variation of $D^{ab}$.

\begin{figure}
\vspace{-1.8cm}
\includegraphics[width=3.2in]{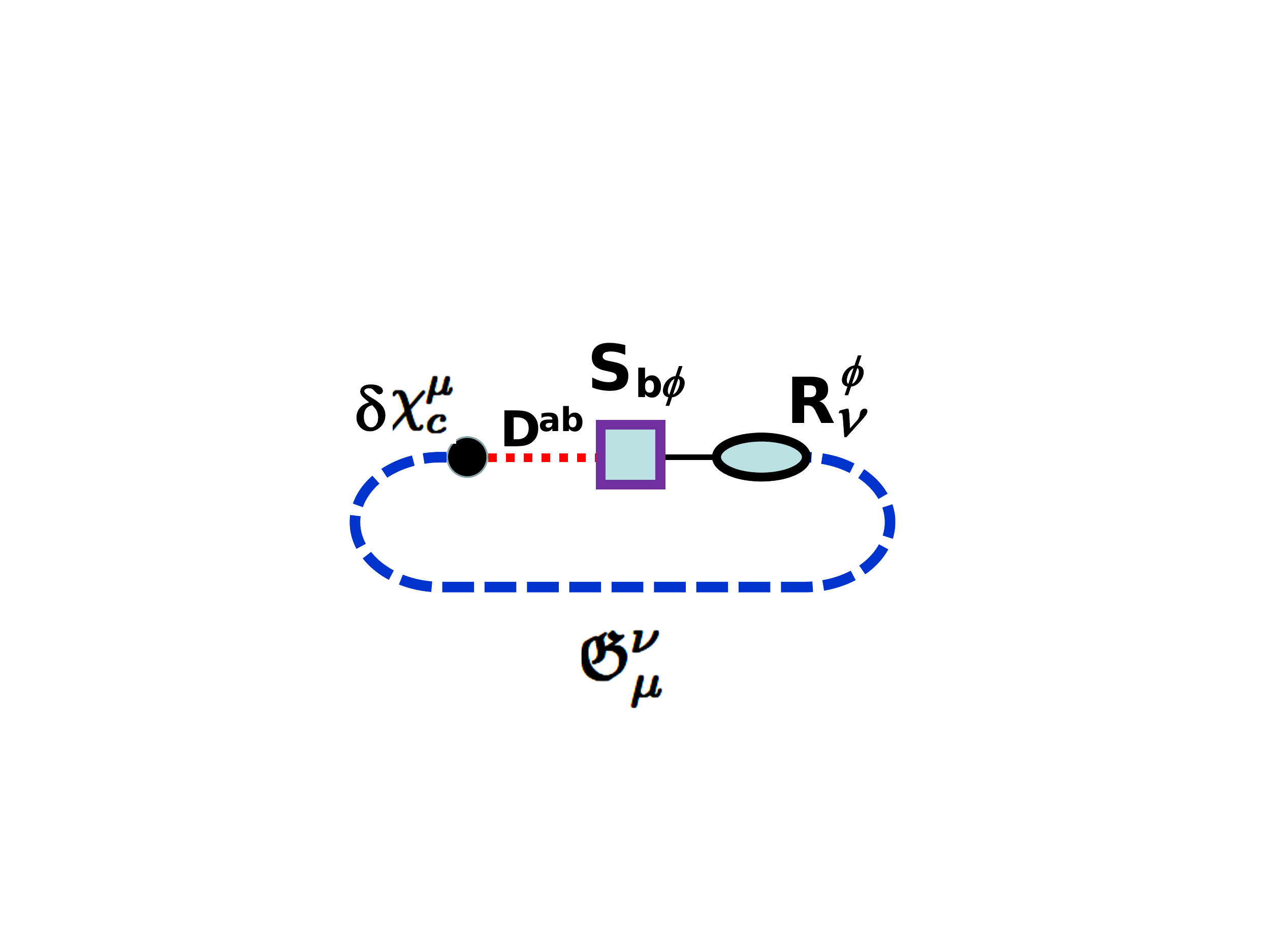}
\vspace{-1.4cm}
\caption{\label{fig:CWLgaugeV} Diagrammatic representation of the term $\delta\chi^\mu_c\,\mathfrak{G}^{\nu}_{\mu}\,D^{cb} S_{b\phi} R^\phi_\nu$ in the integrand of eqtn. (\ref{V}).  The square shows the derivative $\delta^2 S_M/\delta g^b\delta \phi$, the dotted line the graviton propagator $D_{ab}$, and the dashed line the ghost propagator
$\mathfrak{G}^{\nu}_{\mu}$. The 2-point vertex $R^{\phi}_{\nu}$ is shown as an oval. For further explanation see text.  }
\end{figure}

In Fig. \ref{fig:CWLgaugeV} we show the term $(\mathfrak{G}^{\nu}_{\mu}\,D^{cb} S_b\,R^\phi_\nu S_\phi)\,\delta\chi^\mu_c$ which multiplies $e^{-S_M}$ in this expression. We have actually written it slightly differently in the Figure, as $\delta\chi^\mu_c\,\mathfrak{G}^{\nu}_{\mu}\,D^{cb} S_{b\phi} R^\phi_\nu$. To see how one gets this, we rewrite (\ref{d-chiS2}) as follows:
\begin{widetext}
\begin{eqnarray}
\delta_\chi\Big\langle\,\frac{S_{(1)}^2}{2\,l_P^2}\,\Big\rangle_h\,e^{-S_M}
&=& - \mathfrak{G}^{\nu}_{\mu}\,\delta\chi^\mu_c\,D^{cb}
S_b\,\frac\delta{\delta\phi}\Big(R^\phi_\nu\,e^{-S_M}\Big)\;+\;\delta(0)(...)\nonumber\\
    &=& -\frac\delta{\delta\phi}\Big(\,\mathfrak{G}^{\nu}_{\mu}\,\delta\chi^\mu_c\,D^{cb} S_b\,R^\phi_\nu\,e^{-S_M}\Big)\,+\,
    \mathfrak{G}^{\nu}_{\mu}\,\delta\chi^\mu_c\,D^{cb} S_{b\phi}\,R^\phi_\nu\,e^{-S_M}
    \;+\; \delta(0)(...),
    \end{eqnarray}
where in view of the field locality of the generator $\delta
R^\phi_\nu/\delta\phi=\delta(0)\times(...)$, ie., we get power divergent terms which vanish, say, in dimensional regularization. We will disregard these structures here - they are either canceled by the local measure of the path integral, or give rise to anomalies which go beyond this paper. On integration over $\phi$ the total functional derivative term thus disappears, and we have the result shown in Fig. \ref{fig:CWLgaugeV}, viz.,
    \begin{equation}
   \int D\phi\,e^{-S_M}\,\delta_\chi\Big\langle\,
    \frac{S_{(1)}^2}{2\,l_P^2}\,\Big\rangle_h
     \;=\; \int D\phi\,e^{-S_M}\,\big(\,
    \delta\chi^\mu_c\,
    \mathfrak{G}^{\nu}_{\mu}\,D^{cb}
    S_{b\phi} R^\phi_\nu\,\big)\,.         \label{V}
    \end{equation}
 \end{widetext}


Diagrammatically the quantity represents in the Figure is a one-loop object built of the graviton and ghost Green's functions joined via two different local 2-point vertices $\delta\chi^\mu_c$ and $S_{b\phi} R^\phi_\nu$ (and we have shown the latter expanded into the pair of vertices $S_{b\phi}$ and $R^\phi_\nu$, connected by a $\phi\phi$ line).

Remarkably, the gauge variation of the term bilinear in the matter stress tensor, $S_a=T_a/2$, reduces here to a term linear in $S_{b\phi}=\delta^2 S_M/\delta g^b\delta\phi$, which no longer contains any entanglement or path-bunching effect between the different paths or histories, ie., between the different $\phi_i$ in $\varPhi_n=\phi_1,...\phi_n$.

\subsubsection{Loop Terms}
 \label{sec:loop-gauge}

We have seen that the term bilinear in $S_a$ is gauge independent. However to make sure of our results we must also show that the terms linear in $S_a$ are also
gauge independent. As we will see, the functional integration by
parts, of the type just used, will play a critical role in this proof - which actually is non-perturbative in the quantum effects of the matter field.

Our proof is based on checking the gauge variation of each of the 3 other terms, each of which is linear in the gravitational vertices of matter action $S_{(1)}$ and $S_{(2)}$, and each of which contains a gauge loop, in either the graviton or ghost field.

\vspace{3mm}

(i) {\it Seagull Term}: Here we deal with the seagull graph (see Fig. \ref{fig:Qdiagram}(b)). Using the gauge variation of the Green's function, along with the Ward identities derived in the previous sub-section, we get
\begin{widetext}
    \begin{equation}
   \int D\phi\,e^{-S_M}\,\delta_\chi\Big\langle\,
    \frac{S_{(2)}}{l_P^2}\,\Big\rangle_h
   \;=\; \int D\phi\,e^{-S_M}\,
    \delta\chi^\mu_c\,\mathfrak{G}^{\nu}_{\mu}\,D^{cb}
    \Big(S_{b\phi} R^\phi_\nu\,
    +R^a_{\nu,b}S_a \Big)      \label{VI}
    \end{equation}
This quantity has one-loop structure analogous to that just discussed for the path-bunching term; the first of its terms exactly coincides
with (\ref{V}) up to a sign factor, so that they cancel in the gauge variation of the
total sum of terms (\ref{VI}).

\vspace{3mm}

{ii) {\it Ghost Loop Term}: Coming now to the ghost loop term in Fig. \ref{fig:Qdiagram}(c), we see that we have
    \begin{eqnarray}
    \int D\phi\,e^{-S_M}\,\delta_\chi\Big\langle\,
    \frac{S_{(1)}({\rm Tr}\ln\varXi)_{(1)}}{l_P^2}\,
    \Big\rangle_h
    &=&\int D\phi\,e^{-S_M}\,\Big[ S_\phi R^\phi_\alpha\,
    \mathfrak{G}^{\alpha}_{\beta} D^{db}\delta\chi^\beta_d\chi^\mu_c
    R^c_{\nu,b} \nonumber\\
    && \qquad - S_aD^{ab}\,\Big(\delta\chi^\beta_b\, \mathfrak{G}^{\alpha}_{\beta}\,\chi^\mu_c\, R^d_\alpha\, R^c_{\nu,d}-\delta\chi^\mu_c\,R^c_{\nu,b}
    +\chi^\mu_c\,R^c_{\beta,b}\,\mathfrak{G}^{\beta}_{\alpha}\,
    \delta\chi^\alpha_d\,R^d_\nu\Big) \mathfrak{G}^{\nu}_{\mu} \Big] \qquad
    \end{eqnarray}

However, the first contribution is again zero, as in our discussion above, because its integrand is a total functional derivative (modulo a term $R^\phi_{\alpha,\phi}\propto\delta(0)$, which we disregard), since all the factors in $e^{-S_M}S_\phi R^\phi_\alpha\, \mathfrak{G}^{\alpha}_{\beta} D^{db}\delta\chi^\beta_d\chi^\mu_c
    R^c_{\nu,b}\mathfrak{G}^{\nu}_{\mu}$
except $e^{-S_M}S_\phi R^\phi_\alpha$ are $\phi$-independent. Thus we have
    \begin{equation}
    \int D\phi\,e^{-S_M}\,\delta_\chi\Big\langle\,
    \frac{S_{(1)}({\rm Tr}\ln\varXi)_{(1)}}{l_P^2}\,
    \Big\rangle_h= -\int D\phi\,e^{-S_M}\,S_aD^{ab}\,\Big[\,\delta\chi^\beta_b\, \mathfrak{G}^{\alpha}_{\beta}\,\big(R^d_\alpha\, R^c_{\nu,d}\,\chi^\mu_c\big)
    -\delta\chi^\mu_c\,R^c_{\nu,b}
    +\chi^\mu_c\,R^c_{\beta,b}\,\mathfrak{G}^{\beta}_{\alpha}\,
    \delta\chi^\alpha_d\,R^d_\nu\,\Big]
    \,\mathfrak{G}^{\nu}_{\mu} \quad                            \label{VII}
    \end{equation}

\vspace{3mm}

(iii) {\it Graviton Loop Term}: Finally we come to the last term (\ref{IV}), shown in Fig. \ref{fig:Qdiagram}(d). To calculate its variation we note that with a linear gauge, $I_{bcd}=S^G_{bcd}$, and the contraction of this 3-point vertex with the
generator equals
    \begin{eqnarray}
    S^G_{bcd}\,R^d_\mu=-S^G_{bd}\,R^d_{\mu,c}-S^G_{cd}\,R^d_{\mu,b}.
    \end{eqnarray}
Another important contraction can be derived from the Ward identity (\ref{W1}), evaluated on shell so that $S^G_a=0$; it then reads
    \begin{eqnarray}
    S^G_{bd}\,D^{dc}=\delta^c_b-\chi^\alpha_b\,
    \mathfrak{G}^{\beta}_{\alpha}\,R^c_\beta,
    \end{eqnarray}
in which the right hand side is in fact a projector on the non-gauge directions in
configuration space of $g$, $R^b_\mu\big(\,\delta^c_b-\chi^\alpha_b
\mathfrak{G}^{\beta}_{\alpha}R^c_\beta\,\big)=0$.

Using these identities we have
    \begin{eqnarray}
    \int D\phi\,e^{-S_M}\,\delta_\chi\Big\langle\,\frac{I_{(3)}S_{(1)}}{l_P^4}\,
    \Big\rangle_h
    &=&\int D\phi\,e^{-S_M}\Big[\,S_a D^{ab} R^c_{\mu,b}\big(\delta^d_c
    -\chi^\alpha_c \mathfrak{G}^{\beta}_{\alpha}R^d_\beta\big)\,
    \delta\chi^\nu_d\,\mathfrak{G}^{\mu}_{\nu}\nonumber\\
    &&
    +\big(S_b-S_a R^a_\beta \mathfrak{G}^{\beta}_{\alpha}\chi^\alpha_b\big)\,R^b_{\mu,d}\,D^{dc}\,
    \delta\chi^\nu_c\,\,\mathfrak{G}^{\mu}_{\nu}
    +S_a\,D^{ab}\delta\chi^\nu_b\,\mathfrak{G}^{\mu}_{\nu}\,R^c_{\mu,d}\,
    \big(\delta^d_c-\chi^\alpha_c
    \mathfrak{G}^{\beta}_{\alpha}R^d_\beta\big)\,\Big].
    \end{eqnarray}
The second term in the second line here with $R^a_\beta S_a=-R^\phi_\beta S_\phi$ again represents a total derivative in $\phi$ and can be discarded, while in the third line $R^c_{\mu,d}\,\delta^d_c=\delta R^c_\mu/\delta g^c\propto\delta(0)$ and also does not contribute to the final answer. Thus we end up with the expression
    \begin{eqnarray}
    \int D\phi\,e^{-S_M}\,\delta_\chi\Big\langle\,\frac{I_{(3)}S_{(1)}}{l_P^4}\,
    \Big\rangle_h &=&\int D\phi\,e^{-S_M}\Big[\,S_b\,R^b_{\nu,d}\,D^{dc}\,
    \delta\chi^\mu_c\nonumber\\
    &&\quad +\; S_a\,D^{ab}\,\big(R^d_{\nu,b}\, \delta\chi^\mu_d
    -\chi^\alpha_c R^c_{\nu,b}\,\mathfrak{G}^{\beta}_{\alpha}R^d_\beta\, \delta\chi^\mu_d\,\big)
    -S_a\,D^{ab}\delta\chi^\beta_b\,
    \mathfrak{G}^{\alpha}_{\beta}\,\big(R^d_\nu\,R^c_{\alpha,d}\,
    \chi^\mu_c\big)\,\Big]
    \mathfrak{G}^{\nu}_{\mu}  \qquad        \label{VIII}
    \end{eqnarray}
\\}
\end{widetext}

This concludes our results for the variations of each of the 4 graphs in Fig. \ref{fig:Qdiagram}; they are contained in eqtns. (\ref{V}), (\ref{VI}), (\ref{VII}), and (\ref{VIII}). We have seen that we can establish the gauge dependence of each term separately, and establish that the CWL term is itself gauge invariant. It now remains to look at the sum of all 4 terms.

\subsubsection{Gauge Invariance of Total Matter action}
 \label{sec:4terms}

The results contained in eqtns. (\ref{V}), (\ref{VI}), (\ref{VII}) and (\ref{VIII}) are each rather complicated in appearance; and at first glance, there is no particular reason to suppose that their sum should be gauge invariant.

However if we now add them all together, we obtain the rather simple result that
    \begin{eqnarray}
    \delta_\chi\langle\!\langle\, C_n^{(M)} \,\rangle\!\rangle &=& \int
    D\phi\,e^{-S_M}\,S_a\,G^{ab}\delta\chi^\beta_b\,{\cal G}^{\alpha}_{\beta}\, \nonumber
    \\
    && \qquad \times \big(R^d_\alpha\,R^c_{\nu,d}\,-R^d_\nu\,R^c_{\alpha,d}\big)\,
    \chi^\mu_c\,
    {\cal G}^{\nu}_{\mu} \qquad  \label{deltaE0}
    \end{eqnarray}
where $R^d_\alpha\,R^c_{\nu,d}\,-R^d_\nu\,R^c_{\alpha,d}$ represents the commutator of
two diffeomorphism transformations of the gauge field $\Delta^\xi g^a\equiv
R^a_\mu\xi^\mu$ forming the Lie algebra of the general coordinate group, and where $\Delta^\xi
\Delta^\eta-\Delta^\eta \Delta^\xi=\Delta^\zeta$, and
$\zeta^\lambda(x)=\xi^\alpha(x)\partial_\alpha\eta^\lambda(x)
-\eta^\alpha(x)\partial_\alpha\zeta^\lambda(x)$.

Now this commutator can be read as representing the algebra of generators of local gauge transformations; in condensed notation we have
$\zeta^\lambda = {\cal C}^\lambda_{\;\alpha\nu}\xi^\alpha\eta^\nu$, and
    \begin{eqnarray}
    R^d_\alpha\,R^c_{\nu,d}\,-R^d_\nu\,R^c_{\alpha,d}
    = {\cal C}^\lambda_{\;\alpha\nu}\,R^c_\lambda\,.
    \end{eqnarray}
with the structure functions ${\cal C}^\lambda_{\;\alpha\nu}\mapsto
{\cal C}^{\lambda,x}_{\;\alpha,y\;\;\nu,z}\equiv\delta^\lambda_\nu\delta(x,y)
\partial_\alpha\delta(x,z)-(\alpha, y)\leftrightarrow(\nu,z)$.

We now see that the result, for the variation $\delta_\chi\langle\!\langle\, C_n^{(M)} \,\rangle\!\rangle$ in (\ref{deltaE0}) of the matter terms in our effective action, is proportional to the trace of these structure functions. It then follows finally that the variation with respect to gauge conditions of all these matter terms, before the integration over matter fields, gives the result
    \begin{eqnarray}
    \delta_\chi\langle\!\langle\, C_n^{(M)}\,\rangle\!\rangle \;&=& \; \int
    D\phi\,e^{-S_M}\,S_a\,G^{ab}\delta\chi^\beta_b\,
    {\cal G}^{\alpha}_{\beta}\,{\cal C}^\lambda_{\;\alpha\lambda} \nonumber \\
    &\propto& \; \delta(0)\,.
 \label{68}
    \end{eqnarray}
which is proportional to the $\delta(0)$-type term because of the ultra-local nature of the structure functions; from above one has
${\cal C}^\lambda_{\;\alpha\lambda}=\int dx\,{\cal C}^{\lambda,x}_{\;\alpha,y\;\;\lambda,x}\propto \, \delta(0)$. This power divergence vanishes under dimensional regularization; alternatively, it can be canceled by the local measure of the gauge field path integral (which we have disregarded in the foregoing). Thus, finally, we have shown that the set of 4 terms in $C_n^{(M)}$ in (\ref{QnM}) is gauge invariant.

This accomplishes the proof of on-shell gauge independence of the world line correlation term $C_n^{(M)}$ up to the first order of our $l_P^2$-expansion. Beyond this order, the formal proof of this property is based on using a special change of all gravity and matter gauge integration variables under the integration sign \cite{dewitt67b,FP-grav}, and this technique also works in the CWL case.

However the formal implementation of this technique, at different orders in the $l_P^2$-expansion, is much trickier when one includes CWL terms than the way it appears in the conventional loop expansion we have used here. The systematic classification of higher terms and their gauge properties need their own treatment to be considered elsewhere \cite{jordan19}.


\section{Concluding Remarks}
 \label{sec:conclude}


Let us now summarize what we have done here. We can do this from both a mathematical standpoint, and from a physical one.

Mathematically, we have explored the structure of the CWL theory by (i) showing how to do a loop expansion, as well as a perturbative expansion around a background field; and (ii) exhibiting the gauge invariance of the theory. The results show that the CWL theory can be viewed as a legitimate field theory, even though it does violate the quantum mechanical superposition principle. Clearly our formal job is not finished here - for example, we need to investigate the renormalizability of the theory, and look at the structure of perturbation theory at arbitrary order in $l_P^2$.

From a physical standpoint, the CWL theory is in a rather specific sense the
most natural theory one can find in which gravitation is involved in a breakdown of QM. If one asks for a theory in which any gravitational correlations between paths must also satisfy the equivalence principle, then the CWL form follows \cite{stamp15}. The twin requirements of consistent perturbative and classical limits then dictate the ``product CWL" form \cite{BCS18}. 

It then follows that in the CWL framework there are no adjustable parameters, nor any {\it ex cathedra} classical or noise fields - the only fields in the theory are the matter and gravitational fields. There is also no arbitrary distinction between quantum and quantum worlds; one simply passes from one to the other for sufficiently large masses \cite{stamp15}.

Up to order $l_P^2$ we have given a fairly complete characterization here of the theory. The leading departure from conventional quantum gravity (and from standard quantum theory) is given by the path-bunching term, which we have investigated here in detail. To develop the CWL theory into a practical tool, we need to extend our discussion to higher orders in $l_P^2$, and to physically realistic situations.

In work parallel to this, we have succeeded in (i) working out the formal theory of propagators in CWL theory \cite{jordan19} (ii) determining the structure of particle and scalar field propagators to arbitrary orders \cite{JS20} in $l_P^2$, and (iii) calculating the detailed dynamics of single particles and of distributed masses subject to external fields \cite{JYS20}. All of this work is a necessary preliminary to the ultimate goal of the CWL theory, which is to make predictions for the departure from quantum mechanics of the dynamics of objects of mass $\sim M_P$. Viewed from this perspective, the present work consists in laying the theoretical foundations required to do this.


\section{Acknowledgements}
 \label{sec:acknow}


We have benefited from discussions with W.G. Unruh at UBC, with Y. Chen, C. Cheung, and A. Kitaev at Caltech, and with H. Brown and R. Penrose at Oxford. PCES would also like to acknowledge the support and hospitality of Y. Chen, T.F. Rosenbaum, and K.S. Thorne at Caltech. AOB acknowledges the support and hospitality of M. Vessey and Green college at UBC, and of the Peter Wall Institute of Advanced Studies at UBC. 

This work was funded in Canada by the National Science and Engineering Research Council of Canada (NSERC), and by grants from Green College and the Peter wall Institute of Advanced Studies at UBC. In the USA, PCES received support at Caltech from the Simons Foundation (Award 568762) and the National Science Foundation (Award PHY-1733907). The work of AOB was supported in Russia by the RFBR grant No.20-02-00297 and by the Foundation for Theoretical Physics Development ``Basis''.

\end{document}